\documentclass[apjl, iop]{emulateapj}

\slugcomment{Accepted for publication in ApJ}

\usepackage[tight]{subfigure}
\usepackage[colorlinks]{hyperref}
\hypersetup{ colorlinks, linkcolor=blue, citecolor=blue, anchorcolor=blue }
\usepackage{color}

\graphicspath{{images/}}

\newcommand{\Msun}{{\rm M}_{\odot}}
\newcommand{\kms}{km\,s$^{-1}$}
\newcommand{\SiII}{Si~{\sc ii}}
\newcommand{\SII}{S~{\sc ii}}
\newcommand{\NaI}{Na~{\sc i}~D}
\newcommand{\SiIII}{Si~{\sc iii}}
\newcommand{\OI}{O~{\sc i}}
\newcommand{\CII}{C~{\sc ii}}
\newcommand{\FeIII}{Fe~{\sc iii}}
\newcommand{\NiII}{Ni~{\sc ii}}
\newcommand{\FeII}{Fe~{\sc ii}}
\newcommand{\CaII}{Ca~{\sc ii}}
\newcommand{\Nifs}{$^{56}$Ni}

\shorttitle{SN 2011hr}
\shortauthors{Zhang et al.}

\begin{document}

\title{A Luminous Peculiar Type Ia Supernova SN 2011hr: More Like SN 1991T or SN 2007if?}

\author{Ju-Jia Zhang\altaffilmark{1,2,3}, Xiao-Feng Wang\altaffilmark{3},
Michele Sasdelli\altaffilmark{4,5}, Tian-Meng Zhang\altaffilmark{6}, Zheng-Wei
Liu\altaffilmark{7}, Paolo A. Mazzali\altaffilmark{4,5}, Xiang-Cun Meng\altaffilmark{1,2}, Keiichi Maeda\altaffilmark{8,9}, Jun-Cheng Chen\altaffilmark{3}, Fang Huang\altaffilmark{3}, Xu-Lin Zhao\altaffilmark{3}, Kai-Cheng Zhang\altaffilmark{3}, Qian Zhai\altaffilmark{1,2,10}, Elena Pian \altaffilmark{11,12}, Bo Wang\altaffilmark{1,2}, Liang Chang\altaffilmark{1,2}, Wei-Min Yi\altaffilmark{1,2}, Chuan-Jun Wang\altaffilmark{1,2,10}, Xue-Li Wang\altaffilmark{1,2}, Yu-Xin Xin\altaffilmark{1,2}, Jian-Guo Wang\altaffilmark{1,2}, Bao-Li Lun\altaffilmark{1,2},Xiang-Ming Zheng\altaffilmark{1,2}, Xi-Liang Zhang\altaffilmark{1,2}, Yu-Feng Fan\altaffilmark{1,2} and Jin-Ming Bai\altaffilmark{1,2}}
\altaffiltext{1}{Yunnan Observatories (YNAO), Chinese Academy of Sciences, Kunming 650011, China; jujia@ynao.ac.cn}
\altaffiltext{2}{Key Laboratory for the Structure and Evolution of Celestial Objects, Chinese Academy of Sciences, Kunming 650011, China.}
\altaffiltext{3}{Physics Department and Tsinghua Center for Astrophysics (THCA), Tsinghua University, Beijing 100084, China; wang\_xf@mail.tsinghua.edu.cn}
\altaffiltext{4}{Astrophysics Research Institute, Liverpool John
  Moores University, Liverpool Science Park, 146 Brownlow Hill,
  Liverpool L3 5RF, UK}
\altaffiltext{5}{Max-Planck Institute fur Astrophysics, 85748
  Garching, Germany}
\altaffiltext{6}{National Astronomical Observatories of China (NAOC), Chinese Academy of Sciences, Beijing 100012, China.}
\altaffiltext{7}{Argelander-Institut f\"{u}r Astronomie, Auf dem H\"{u}gel 71, D-53121, Bonn, Germany.}
\altaffiltext{8}{Department of Astronomy, Kyoto University, Kyoto, 606-8502, Japan}
\altaffiltext{9}{Kavli Institute for the Physics and Mathematics of the Universe (WPI), University of Tokyo, 5-1-5 Kashiwanoha, Kashiwa, Chiba 277-8583, Japan}
\altaffiltext{10}{University of Chinese Academy of Sciences, Chinese Academy of Sciences, Beijing 100049, China.}
\altaffiltext{11}{INAF-IASF-Bo, via Gobetti, 101, I-40129 Bologna, Italy.}
\altaffiltext{12}{Scuola Normale Superiore, Piazza dei Cavalieri 7, 56126 Pisa, Italy.}

\begin{abstract}
Photometric and spectroscopic observations of a slowly declining, luminous Type Ia supernova (SN Ia) SN  2011hr in the starburst galaxy NGC 2691 are presented.  SN~2011hr is found to peak at $M_{B}=-19.84 \pm 0.40\,\rm{mag}$, with a post-maximum decline rate $\Delta$m$_{15}$(B) = 0.92 $\pm$ 0.03\,$\rm{mag}$. From the maximum-light bolometric luminosity, $L=(2.30 \pm 0.90) \times 10^{43}\,\rm{erg\,s^{-1}}$, we estimate the mass of synthesized \Nifs\ in SN~2011hr to be $M(\rm{^{56}Ni})=1.11 \pm 0.43\,M_{\sun}$. SN 2011hr appears more luminous than SN 1991T at around maximum light, and the absorption features from its intermediate-mass elements (IMEs) are noticeably weaker than the latter at similar phases. Spectral modeling suggests that SN 2011hr has the IMEs of $\sim$\,0.07 M$_{\sun}$ in the outer ejecta, which is much lower than the typical value of normal SNe Ia (i.e., 0.3 -- 0.4 M$_{\sun}$) and is also lower than the value of SN 1991T (i.e., $\sim$\,0.18 M$_{\sun}$). These results indicate that SN~2011hr may arise from a Chandrasekhar-mass white dwarf progenitor that experienced a more efficient burning process in the explosion. Nevertheless, it is still possible that SN~2011hr may serve as a transitional object connecting the SN 1991T-like SNe Ia with the superluminous subclass like SN 2007if given that the latter also shows very weak IMEs at all phases.

\end{abstract}


\section{Introduction}
\label{sect:Intro}

Type Ia supernovae (SNe Ia), one of the most luminous stellar explosion, could explode within a similar mechanism: complete explosive destruction of  carbon-oxygen white dwarf (C-O WD) reaching the Chandrasekhar mass limit (i.e., $\sim$\,1.4 M$_{\sun}$ for the C-O WD without rotation), most likely via accretion in a binary system, although other channels have been proposed (see e.g., the review by \citealp{Maoz14}). They show strikingly similar photometric and spectroscopic behavior in optical (i.e., \citealp{Sun96,Filip97}), and the remaining scatter can be better understood in terms of an empirical relation between light-curve width (i.e., $\Delta$m$_{15}$(B)) and luminosity (i.e., Width-Luminosity Relation [WLR];  \citealp{Phillip93}). Based on that relation, SNe Ia have been successfully applied as luminosity distance indicators for measuring the expansion history of the universe \citep{Riess98,Schmidt98,Perl99}.

Observationally, SNe Ia could be divided into several different subclasses according to their spectral characteristics \citep{Ben05, Branch09, Wang09a}. For example, SN 1991T \citep{Filip92a,Phillip92} is spectroscopically different from the bulk of SNe\,Ia, lying at the luminous end of luminosity distribution. Its featureless spectra at early phase are characterized by appearances of strong \FeIII~lines and weak or absent lines of intermediate-mass elements (IMEs, e.g., \SiII, \SII\ and \CaII) which are usually strong in the spectra of normal SNe Ia. It had a slowly declining light curve (e.g., $\Delta m_{15}$(B) = 0.94, \citealp{Lira98}) and is $\sim$\,0.6 mag more luminous than normal ones \citep{Filip92a}. However, \citet{Gibson} suggested that SN\,1991T is indistinguishable from normal SNe\,Ia in luminosity. \citet{Altavilla} also showed that SN 1991T fits in the WLR and that its peculiarities are only spectroscopic. Through modeling of a series of early spectra, \citet{Paolo95} found that the absence of \FeII, \SiII, and \CaII\ lines and the presence of strong \FeIII\ lines are due to the low abundance of IMEs and the high ionization caused by the high luminosity. Superluminous SNe~Ia such as 2003fg \citep{Howell06}, 2006gz \citep{Hicken07}, 2007if \citep{Scalzo07if} and 2009dc \citep{Silver11,Tau09dc,Hach09dc} are called Super-Chandrasekhar (SC) SN\,Ia candidates. These SNe~Ia are required to have progenitor masses exceeding the Chandrasekhar-mass limit if their luminosities are entirely attributed to the radioactive decay of \Nifs.

In this paper, we present observations of SN 2011hr (R.A.\,=$\,08^{\rm h}\,54^{\rm m}\,46^{\rm s}$.03, Decl.\,=\,$+39^\circ$\,32\arcmin\,16.1\arcsec) which was discovered on UT Nov. 8.54 2011 (UT time is used throughout this paper) by the Katzman Automatic Imaging Telescope, Lick Observatory, USA \citep{CBET2901a}. The supernova is located 3\arcsec.4~west and 3\arcsec.7 north of the centre of NGC 2691, which is equivalent to a projected distance $\sim$\,1.5 kpc using the distance of the host adopted in this paper (i.e., D\,$\approx$\,60.0 Mpc). Two days after the discovery, an optical spectrum was taken with the Li-Jiang 0.8 m telescope (hereafter is LJT) of Yunnan observatories (YNAO), China, showing that it was a young SN\,1991T-like SN Ia at about ten days before the maximum light, with prominent iron lines in the spectrum \citep{CBET2901b}. A follow-up photometric and spectroscopic observing campaign was immediately established for SN\,2011hr.

This paper is organised as follows. Photometric observations and data reductions are described in Section \ref{sect:Ph}, where the light and color curves are investigated. Section \ref{sect:Sp} presents the spectroscopic observations. In Section \ref{sect:Discu} we derive the bolometric luminosity and the mass of \Nifs\ synthesized in the explosion, and model the early-time spectra using abundance tomography. Moreover, possible explosion mechanism is also discussed in this section. A brief summary is given in Section \ref{sect:con}.
 \begin{figure}
\centering
\includegraphics[width=8.5cm,angle=0]{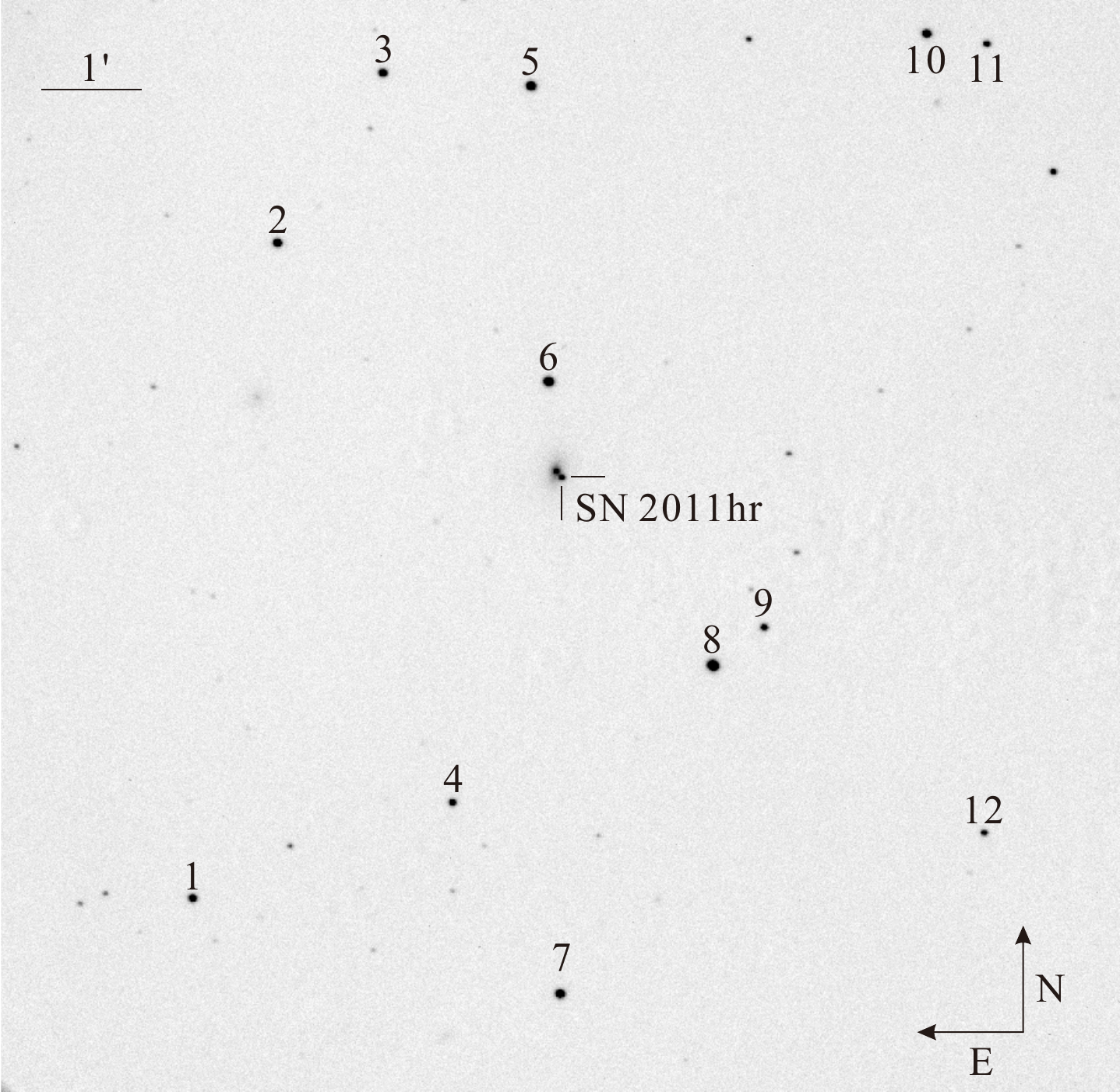}
 \caption{The $R$ band image of SN 2011hr, which was taken on Nov. 14.81 2011 with the Tsinghua-NAOC 0.8 m telescope. The mean FWHM of this image is $2\arcsec.60$ under the scale of $0\arcsec.52$/pixel. The supernova and reference stars are marked. North is up, and east is to the left.}
\label{<img>}
\end{figure}

\section{Photometry}
\label{sect:Ph}
\subsection{Observations and Data Reduction}
The photometric observations of SN 2011hr were performed in broad $UBVRI$ bands with the Tsinghua-NAOC 0.8 m telescope (hereinafter TNT, \citealp{wang08,TNT}) at Xing-Long Observatory of NAOC. This telescope is equipped with a 1340\,$\times$\,1300 CCD (PI VersArray: 1300B), providing a field-of-view (FOV) of $11\arcmin.5\,\times\,11\arcmin.2$ with a spatial resolution of $\sim$\,0$\arcsec$.52 pixel$^{-1}$. The typical exposure times are 180s in the BVRI bands and 300s in the U band when taking images of SN 2011hr. These images have typical full-width-half-maximum (FWHM) of about 2$\arcsec$.5. The CCD images were reduced using standard IRAF routines. The data collections started from $t\approx-14$ days and lasted until $t\approx+70$ days relative to the $B$ band maximum light. All CCD images were reduced using the IRAF\footnote{IRAF, the Image Reduction and Analysis Facility, is  distributed by the National Optical Astronomy Observatory, which is operated by the Association of Universities for Research in Astronomy(AURA), Inc. under cooperative agreement with the National Science Foundation(NSF).} standard  procedure.

\begin{figure*}
\centering
\includegraphics[width=15.4cm,angle=0]{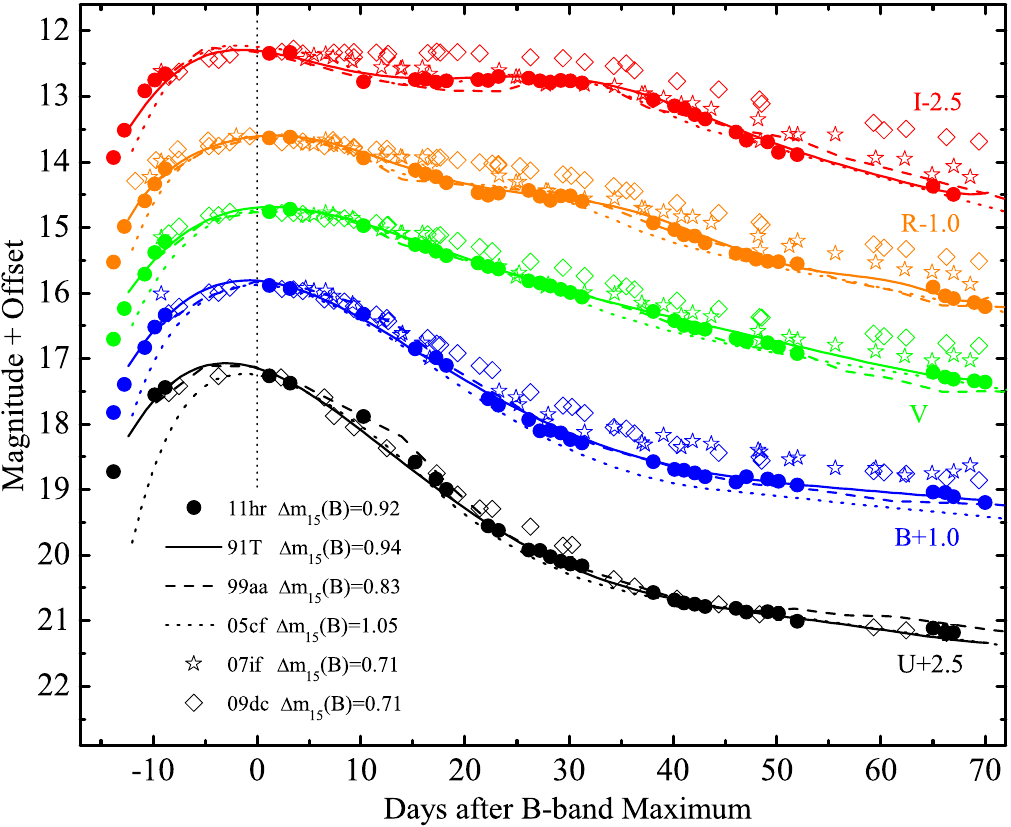}
 \caption{The $UBVRI$ band light curves of SN 2011hr, compared with those of SNe 1991T, 1999aa, 2005cf, 2007if, and 2009dc ( see text for references). The light curves of SN\,2011hr are shifted vertically by numbers as indicated in the plot and those of the comparison SNe are normalized to the peak magnitudes of SN\,2011hr. The light curves of SNe\,1991T, 1999aa, and 2005cf have been slightly smoothed with a low-order polynomial for a better display.}
\label{<LC>}
\end{figure*}

Since SN\,2011hr exploded near the center of its host galaxy, the step of subtracting the galaxy template from the SN images is necessary for accurate photometry, in particular when the SN became faint and/or the images were taken under poor seeing conditions. Higher quality template images were obtained with the TNT on May 11, 2013, when the SN was sufficiently faint.
Using the galaxy-subtracted images, we measured the instrumental magnitudes of the SN and the local standard stars (labeled in Figure \ref{<img>}) by aperture photometry, performed using the IRAF DAOPHOT package (Stetson 1987). The radius for the photometric aperture is set as 2.0 times the FWHM; the sky background light was determined from the median counts at a radius that is 5.0 times the FWHM. Twelve local standard stars in the field of SN 2011hr are labeled in Figure \ref{<img>}. We transformed the instrumental magnitudes of these reference stars to the standard Johnson $UBV$ \citep{Johnson} and Kron-Cousins $RI$ \citep{Cousins} systems through the following transforming equations \citep{TNT}:

\begin{equation}
U = u + 0.22(U-B),
\label{eq:1}
\end{equation}
\begin{equation}
B = b + 0.15(B-V),~
\label{eq:2}
\end{equation}
\begin{equation}
V = v - 0.06(B-V),~
\label{eq:3}
\end{equation}
\begin{equation}
R = r - 0.10(V-R),~~
\label{eq:4}
\end{equation}
\begin{equation}
I = i + 0.02(V-I) .~~~~
\label{eq:5}
\end{equation}
where $UBVRI$ represent the magnitudes in the standard system and $ubvri$ represent the instrumental magnitudes. The color term coefficient was determined by observing \citet{Landolt} standards on photometric nights. The magnitudes of these reference stars are listed in Table \ref{Tab:Photo_stand}, and they are then used to convert the photometry of SN\,2011hr to the standard $UBVRI$ magnitudes as shown in Table 2.

\begin{deluxetable}{cccccc}[!th]
\tablecaption{Magnitudes of the Photometric Standards in the Field of SN 2011hr}
\tablewidth{0pt}
\tablehead{\colhead{Star} &  \colhead{$U$(mag)} & \colhead{$B$(mag)} & \colhead{$V$(mag)} & \colhead{$R$(mag)} & \colhead{$I$(mag)} }
\startdata
1 & 16.01(03) & 15.27(02) & 14.50(03) & 14.00(02) & 13.66(02) \\
2 & 14.15(02) & 14.15(02) & 13.78(02) & 13.48(02) & 13.32(01) \\
3 & 17.03(04) & 15.66(02) & 14.58(03) & 13.79(02) & 13.33(01) \\
4 & 17.99(04) & 16.61(03) & 15.31(03) & 14.39(02) & 13.72(02) \\
5 & 16.06(03) & 14.81(02) & 13.88(02) & 13.22(02) & 12.88(01) \\
6 & 16.27(03) & 14.96(03) & 13.71(02) & 12.99(02) & 12.46(01) \\
7 & 14.36(03) & 14.26(02) & 13.67(02) & 13.35(02) & 13.17(02) \\
8 & 13.94(02) & 13.47(01) & 12.76(01) & 12.40(01) & 12.15(01) \\
9 & 15.91(03) & 15.53(02) & 14.91(03) & 14.52(02) & 14.26(02) \\
10 & 16.57(04) & 15.67(03) & 14.44(03) & 13.48(03) & 12.76(03) \\
11 & 17.30(05) & 16.28(04) & 15.38(04) & 14.70(04) & 14.39(04) \\
12 & 16.41(03) & 16.13(02) & 15.39(03) & 14.96(03) & 14.67(02)
\enddata
\tablecomments{Uncertainties, in units of 0.01 mag, are 1 $\sigma$. Figure 1 shows a finding chart of SN 2011hr and the comparison stars.}
\label{Tab:Photo_stand}

\end{deluxetable}

\subsection{Light Curves}
\label{subsect:OPLC}

\begin{deluxetable*}{ccccccc}
\tablewidth{0pt}
\tablecaption{The $UBVRI$ Photometry of SN 2011hr}

\tablehead{\colhead{MJD} & \colhead{Epoch\tablenotemark{a}} & \colhead{U(mag)} & \colhead{B(mag)} & \colhead{V(mag)} & \colhead{R(mag)} & \colhead{I(mag)}}

\startdata

55874.04    &   -15.08  &    \nodata    &   \nodata     &   \nodata     &   16.90(20)   &    \nodata   \\
55875.08    &   -14.04  &    \nodata    &   \nodata     &   \nodata     &   17.10(20)   &    \nodata    \\
55875.34	&	-13.78	&	16.17(08)	&	16.90(03)	&	16.71(02)	&	16.53(02)	&	16.43(02)	\\
55876.36	&	-12.76	&	\nodata   	&	16.47(02)	&	16.24(01)	&	15.98(02)	&	16.01(01)	\\
55878.32	&	-10.80	&	\nodata	    &	15.91(02)	&	15.71(04)	&	15.59(02)	&	15.41(01)	\\
55879.30	&	-9.82	&	15.00(05)	&	15.60(02)	&	15.38(02)	&	15.33(01)	&	15.24(01)	\\
55880.31	&	-8.81	&	14.89(05)	&	15.41(02)	&	15.22(01)	&	15.10(01)	&	15.15(03)	\\
55890.31	&	1.19	&	14.71(03)	&	14.97(02)	&	14.76(01)	&	14.63(01)	&	14.84(01)	\\
55892.31	&	3.19	&	14.82(03)	&	15.01(02)	&	14.73(01)	&	14.62(01)	&	14.83(01)	\\
55899.36	&	10.24	&	15.33(04)	&	15.40(02)	&	14.97(05)	&	14.93(01)	&	15.27(04)	\\
55904.33	&	15.21	&	16.03(08)	&	15.93(02)	&	15.26(01)	&	15.12(04)	&	15.24(03)	\\
55905.33	&	16.21	&	\nodata	&	\nodata	&	15.29(05)	&	15.18(04)	&	15.23(05)	\\
55906.31	&	17.19	&	16.28(05)	&	16.06(03)	&	15.36(02)	&	15.22(03)	&	15.28(01)	\\
55907.33	&	18.21	&	16.45(06)	&	16.18(04)	&	15.43(01)	&	15.32(01)	&	15.26(01)	\\
55910.35	&	21.23	&	\nodata	&	\nodata	&	15.54(04)	&	15.47(06)	&	15.24(05)	\\
55911.33	&	22.21	&	17.00(05)	&	16.69(02)	&	15.59(05)	&	15.51(04)	&	15.25(04)	\\
55912.33	&	23.21	&	17.07(05)	&	16.79(05)	&	15.63(01)	&	15.47(01)	&	15.19(03)	\\
55915.26	&	26.14	&	17.37(05)	&	17.01(03)	&	15.81(01)	&	15.43(01)	&	15.22(04)	\\
55916.31	&	27.19	&	17.38(05)	&	17.18(02)	&	15.84(05)	&	15.52(02)	&	15.26(02)	\\
55917.33	&	28.21	&	17.47(09)	&	17.18(02)	&	15.89(05)	&	15.59(01)	&	15.29(03)	\\
55918.31	&	29.19	&	17.54(06)	&	17.21(07)	&	15.94(01)	&	15.51(01)	&	15.26(04)	\\
55919.25	&	30.13	&	17.58(10)	&	17.31(02)	&	15.99(01)	&	15.51(01)	&	15.26(03)	\\
55920.39	&	31.27	&	17.61(05)	&	17.37(01)	&	16.06(05)	&	15.60(01)	&	15.29(02)	\\
55927.23	&	38.11	&	18.02(06)	&	17.65(07)	&	16.28(05)	&	15.93(05)	&	15.55(05)	\\
55929.24	&	40.12	&	18.13(06)	&	17.77(08)	&	16.42(05)	&	16.03(05)	&	15.64(06)	\\
55930.17	&	41.05	&	18.18(10)	&	17.78(04)	&	16.49(05)	&	16.11(04)	&	15.68(08)	\\
55931.21	&	42.09	&	18.20(09)	&	17.82(04)	&	16.53(04)	&	16.13(02)	&	15.78(02)	\\
55932.21	&	43.09	&	18.23(06)	&	17.88(05)	&	16.55(04)	&	16.23(01)	&	15.84(03)	\\
55935.19	&	46.07	&	18.26(07)	&	17.96(04)	&	16.69(04)	&	16.39(01)	&	16.04(02)	\\
55936.20	&	47.08	&	18.32(05)	&	17.88(03)	&	16.74(05)	&	16.42(04)	&	16.17(05)	\\
55937.11	&	47.99	&	\nodata	&	\nodata	&	\nodata	&	16.48(03)	&	\nodata	\\
55938.21	&	49.09	&	18.31(09)	&	17.92(04)	&	16.76(05)	&	16.51(05)	&	16.19(03)	\\
55939.25	&	50.13	&	18.34(10)	&	17.95(03)	&	16.82(04)	&	16.52(08)	&	16.35(02)	\\
55941.08	&	51.96	&	18.46(08)	&	18.01(03)	&	16.92(05)	&	16.55(03)	&	16.39(06)	\\
55954.12	&	65.00	&	18.56(10)	&	18.11(03)	&	17.21(02)	&	16.91(05)	&	16.87(05)	\\
55955.30	&	66.18	&	18.62(12)	&	18.12(07)	&	17.28(04)	&	17.04(05)	&	\nodata	\\
55956.08	&	66.96	&	18.63(12)	&	18.19(08)	&	17.31(05)	&	17.08(05)	&	16.99(07)	\\
55958.06	&	68.94	&	\nodata	&	\nodata	&	17.34(07)	&	17.14(06)	&	\nodata	\\
55959.10	&	69.98	&	\nodata	&	18.27(08)	&	17.36(07)	&	17.21(08)	&	\nodata	
\enddata
\tablecomments{Uncertainties, in units of 0.01 mag, are 1$\sigma$; MJD = JD $-$ 2400000.5.}
\tablenotetext{a}{Relative to the epoch of $B$ band maximum on Nov. 23.49 2011, JD. 2455889.62.}
\label{Tab:Photo}
\end{deluxetable*}


Figure \ref{<LC>} displays the $UBVRI$ band light curves of SN\,2011hr (see details in Table \ref{Tab:Photo}). Some basic photometric parameters such as the maximum-light dates, peak magnitudes and magnitude decline rates ($\Delta$m$_{15}$, \citealp{Phillip93}) can be derived from a polynomial fit to the observed light curves, which are listed in Table \ref{Tab:LV_par}.  We find that SN 2011hr reached a $B$ band maximum brightness of $14.95\,\pm\,0.03$ mag on JD 2455889.62\,$\pm$\,0.30 (Nov. 23.12 2011), with a decline rate as $\Delta m_{15}$(B)\,=\,0.92\,$\pm$\,0.03 mag. The slower decline rate suggests that SN 2011hr should be intrinsically luminous if it follows the WLR (e.g., \citealp{Phillip93,Riess96,Goldhaber01,Guy05}), see also Table \ref{Tab:LV_par} and detailed discussion in Section \ref{subsect:Dist}. After correcting for a Galactic reddening of E$(\bv)$\,=\,0.02 mag \citep{Schle98}, the color index B$_{max}$\,$-$\,V$_{max}$ becomes 0.23\,$\pm$\,0.04 mag. On the other hand, the corresponding color index is about 0.13\,$\pm$\,0.03 mag for SN 1991T. This suggests both SN 2011hr and SN 1991T suffer a significant host-galaxy reddening assuming that their intrinsic colors are comparable to that established for normal SNe~Ia (e.g., \citealp{Phillip99, Wang09b}), see also section \ref{subsect:CC} and \ref{subsect:red}.

Overplotted in Figure \ref{<LC>} includes the light curves of other well-observed SNe including overluminous SN Ia 1991T ($\Delta$m$_{15}$\,=\,0.94 mag; \citealt{91T_14} and therein), SN 1999aa ($\Delta$m$_{15}$(B)\,=\,0.83 mag; \citealp{Jha}), the `golden standard' SN Ia 2005cf ($\Delta$m$_{15}$(B)\,=\,1.05\,mag; \citealp{Wang09b}), the superluminous SN 2007if ($\Delta$m$_{15}$(B)\,=\,0.71; \citealp{Scalzo07if}) and SN 2009dc($\Delta$m$_{15}$(B)\,=\,0.71; \citealp{Tau09dc}). In general, SN 2011hr show close resemblances to SN 1991T in $UBVRI-$band photometry. Compared to SN 2011hr/SN 1991T, the superluminous object SN 2009dc shows apparently broader light curves, especially in the redder bands where its secondary maximum/shoulder feature is more prominent. Note that another so-called superluminous type Ia SN 2007if is found to be more similar in morphology to SN 2011hr/SN 1991T.

\begin{deluxetable}{cccccc}[]
\tablewidth{0pt}
\tablecaption{Light Curve Parameters of SN 2011hr}

\tablehead{\colhead{Band} & \colhead{$\lambda_{mid}$} & \colhead{t$_{max}$\tablenotemark{a}} & \colhead{m$_{peak}$\tablenotemark{b}} & \colhead{$\Delta$ m$_{15}$\tablenotemark{b}}& \colhead{M$_{peak}$\tablenotemark{b}}  \\
\colhead{} & \colhead{(\AA)} & \colhead{(-2455000.5)} & \colhead{(mag)} & \colhead{(mag)} & \colhead{(mag)}  }

\startdata
U & 3650 & 887.39(50) & 14.63(03) & 1.04(05) &-20.34(45) \\
B & 4450 & 889.12(30) & 14.95(02) & 0.92(03) &-19.84(40)\\
V & 5500 & 889.95(30) & 14.70(03) & 0.59(03) &-19.87(35)\\
R & 6450 & 890.01(30) & 14.62(02) & 0.53(05) &-19.81(30)\\
I & 7870 & 890.10(40) & 14.70(03) & 0.45(05) &-19.57(30)
\enddata

\tablenotetext{a}{1 $\sigma$ uncertainties, in units of 0.01 day.}
\tablenotetext{b}{1 $\sigma$ uncertainties, in units of 0.01 mag.}

\label{Tab:LV_par}
\end{deluxetable}

\begin{figure*}[!th]
\centering
\includegraphics[width=17cm,angle=0]{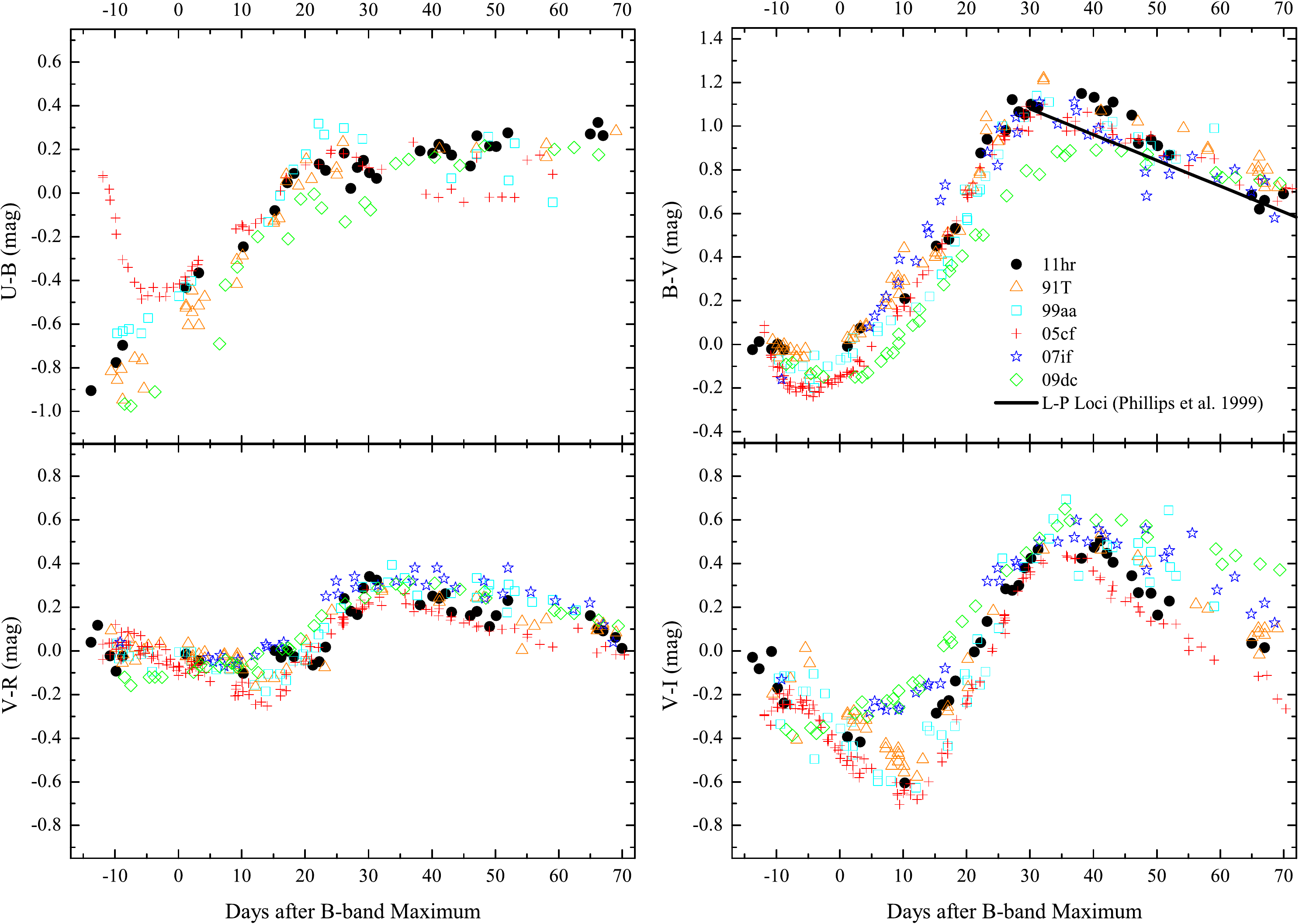}
 \caption{Optical color curves of SN 2011hr, in comparison with those of SN 1991T, SN 1999aa, SN2005cf, SN 2007if, and SN 2009dc. See text for details.}
\label{<CC>}
\end{figure*}

\subsection{Color Curves}
\label{subsect:CC}

Figure \ref{<CC>} shows the color curves of SN 2011hr. Over-plotted are the color curves of SN 1991T, SN 1999aa, SN 2005cf, SN 2007if and SN 2009dc.  All of these color curves (including those of SN 2011hr) are corrected for the reddening of the Milky Way and the host galaxy. The reddening of SN 2011hr is discussed in the following section.

At early time, the $U\,-\,B$ colors of the luminous/superluminous SNe Ia (including SN 2011hr) show distinct behavior compared to the normal SN 2005cf.  The former group tends to become progressively redder after explosion while SN 2005cf shows a `red-blue-red' evolutionary trend at similar phase. At around the maximum light, the $B - V$ color of SN 2011hr and SN 1991T is redder than that of other comparison SNe Ia. SN 2011hr and SN 1991T are also found to be redder by 0.1-0.2 mag than that predicted by the $Lira-Phillips$ relation (\citealp{Phillip99}; see solid-line in Figure \ref{<CC>}). Among our comparison sample, SN 2009dc shows the bluest $U - B$ and $B - V$ colors at t $<$ +35 days while it (and SN 2007if) shows redder $V - R$ and $V - I$ colors. Large scatter can be seen in the $V - I$ color where the strength of \CaII\ NIR triplet could have a dominant effect. In general, SN 2011hr shows color evolution that is quite similar to SN 1991T and SN 2007if at all phases.

\subsection{Extinction}
\label{subsect:red}
The Milky Way reddening of SN 2011hr is $E(\bv)_{\rm Gal}$ = 0.02 mag \citep{Schle98}, corresponding to an extinction of 0.06 mag adopting the standard Galactic reddening law (i.e., $R_V$\,=\,3.1; \citealp{Cardelli}). The reddening due to the host galaxy can be estimated by several empirical methods. For instance, a comparison of the late-time $B - V$ color with that predicted by the Lira-Phillips relation \citep{Phillip99} gives a host-galaxy reddening of $E(B - V)_{host}$\,=\,0.29$\pm$0.07 mag for SN 2011hr. The peak B$_{max}$ $-$ V$_{max}$ color has also been used as a better reddening indicator for SNe Ia \citep{Phillip99,Wang09b}, based on its dependence on $\Delta$m$_{15}$(B). For SN 2011hr, the observed peak B $-$ V color is 0.23\,$\pm$\,0.04 mag, which suggests a host-galaxy reddening of E(B-V) = 0.32\,$\pm$\,0.06 mag. However, a scatter of about 0.1 mag might exist for the above estimates considering that such overluminous SNe Ia may have intrinsically different colors relative to the normal SNe Ia.

The low-resolution spectra of SN 2011hr show prominent signatures of \NaI\ absorption, for which the equivalent width (EW) of this absorption can be derived by using double-Gaussian fit as shown in Figure \ref{<Whsp>}). The best-fit yields $EW$(\NaI) = 1.42 $\pm\,0.02$ \AA\, with $EW_{\rm D1}$ = 0.84\,$\pm\,0.01$\,\AA\ and $EW_{\rm D2}$ = 0.58\,$\pm\,0.01$\,\AA\, respectively. Based on an empirical correlation between reddening and $EW$ of Na~{\sc i}\,D, namely, $E(B-V) = 0.16\,EW - 0.01$ [\AA] \citep{Tura16}, we estimate  $E(B - V)_{\rm host}$ to be 0.22 $\pm$ 0.10 mag for SN 2011hr, which is consistent the result derived from the photometric $B - V$ color within 1\,$\sigma$ error.

Combining the results from both photometry and \NaI\ absorption, a host-galaxy reddening of $E(B - V)_{\rm host}$ = 0.25\,$\pm$\,0.10 can be obtained for SN 2011hr. On the other hand, the spectral modeling presented in section \ref{subsect:Spmo} suggests a moderate host-galaxy reddening, i.e., $E(B - V)_{\rm host}$ = 0.15 mag, which seems to be more reasonable as it matches the observed colors. Therefore, a mean value $E(B - V)_{\rm total} = 0.22\,\pm\,0.05$ mag is adopted as the total reddening of SN 2011hr in this paper.

\subsection{Distance}
\label{subsect:Dist}
The actual luminosity of SN 2011hr is affected by the uncertainty in the distance to the SN. Table \ref{Tab:disct} lists the measurements obtained using different methods. One can see that these measurements show significant dispersion. For example, the distance derived from the redshift of the host galaxy NGC 2691 and the Hubble flow is 59.3\,$\pm$\,4.0 Mpc \citep{Mould00}, which agrees with the result from the Tully-Fisher relation (60.2\,$\pm$\,10.0 Mpc; \citealp{DisTF}). These two values are larger than the luminosity distance derived for the SN using the original \citet{Phillip93} relation but is close to that derived from the modified relation for 91T-like events in \citet{Blondin12}, as listed in Table \ref{Tab:disct}.

On the other hand, we can also estimate the distance to SN 2011hr based on its near-infrared (NIR) magnitude, since the $JHK-$band luminosities of SNe Ia are relatively insensitive to the reddening and are more uniform around the peak compared to the corresponding values in the optical bands \citep{Meik00,Kri04,Barone-Nugent12}. Additionally, some peculiar SNe Ia like SNe 1999aa, 1999ac, and 1999aw are found to have normal luminosities in the NIR bands \citep{Kri04}, though they appear somewhat peculiar in  optical bands. Based on the NIR photometry of 13 SNe Ia (including SN 2011hr), \citet{Anja11hr} estimated their mean absolute peak magnitude as $-18.31 \pm 0.02$ mag in the $H$ band. Taking the observed $H$ band peak magnitude of 15.02\,$\pm\,0.20$ mag for SN 2011hr \citep{Anja11hr} and assuming that it has a similar absolute $H$ band magnitude, we  get a distance modulus $m - M = 33.20\,\pm\,$0.20 mag or $D = 43.7\,\pm\,$5.0 Mpc. This estimate is much smaller than the Tully-Fisher distance and that inferred from the Hubble flow. Note, however, that the dispersion of the sample from \citet{Anja11hr} will decrease significantly from $\sigma$$_{\rm H}$ = 0.23 mag to $\sim$ 0.18 mag if SN 2011hr was excluded in the analysis. This is due to that SN 2011hr is actually more luminous than normal SNe Ia by $\sim$ 0.7 mag in the $H$ band.

\begin{deluxetable}{lcc}
\tablecaption{Distances to SN\,2011hr and NGC\,2691}
\tablehead{\colhead{Method} & \colhead{Details} & \colhead{Results (Mpc)} }
\startdata
Hubble Flow\tablenotemark{a}  & 4269 \kms & 59.3$\pm$4\\
Tully-Fisher\tablenotemark{b}  & $JHK$ bands & 60.2$\pm$10 \\
Phillips  Relation\tablenotemark{c}  & $E(B-V)$=0.17 & 48.24$\pm$8 \\
Phillips  Relation\tablenotemark{d}  & $E(B-V)$=0.17 & 56.58$\pm$4 \\
NIR luminous\tablenotemark{e} & $H$ band & 43.7$\pm$5
\enddata
\tablenotetext{a}{Corrected for Virgo infall, GA and Shapley \citep{Mould00}, on the scale of $H_0$=72 \kms Mpc$^{-1}$}
\tablenotetext{b}{Mean distance from estimates in the JHK bands \citep{DisTF,Tully}}
\tablenotetext{c}{$\Delta$m$_{15}$(B)=0.92; \citet{Phillip93}}
\tablenotetext{d}{Modified \citet{Phillip93} relation in the Fig. 13 of \citet{Blondin12} for 91T-like events}
\tablenotetext{e}{\citet{Anja11hr}}
\label{Tab:disct}
\end{deluxetable}

As the distance derived from Hubble expansion is independent of corrections for reddening and agrees well with the estimate from the Tully-Fisher relation, a distance of D = 60.0\,$\pm$\,6.0 Mpc is thus adopted for SN 2011hr in this paper.

\section{Spectroscopy}
\label{sect:Sp}

\subsection{Observations and Data Reduction}
\label{subsect:ObsofSP}

\begin{figure}[]
\centering
\includegraphics[width=8.5cm,angle=0]{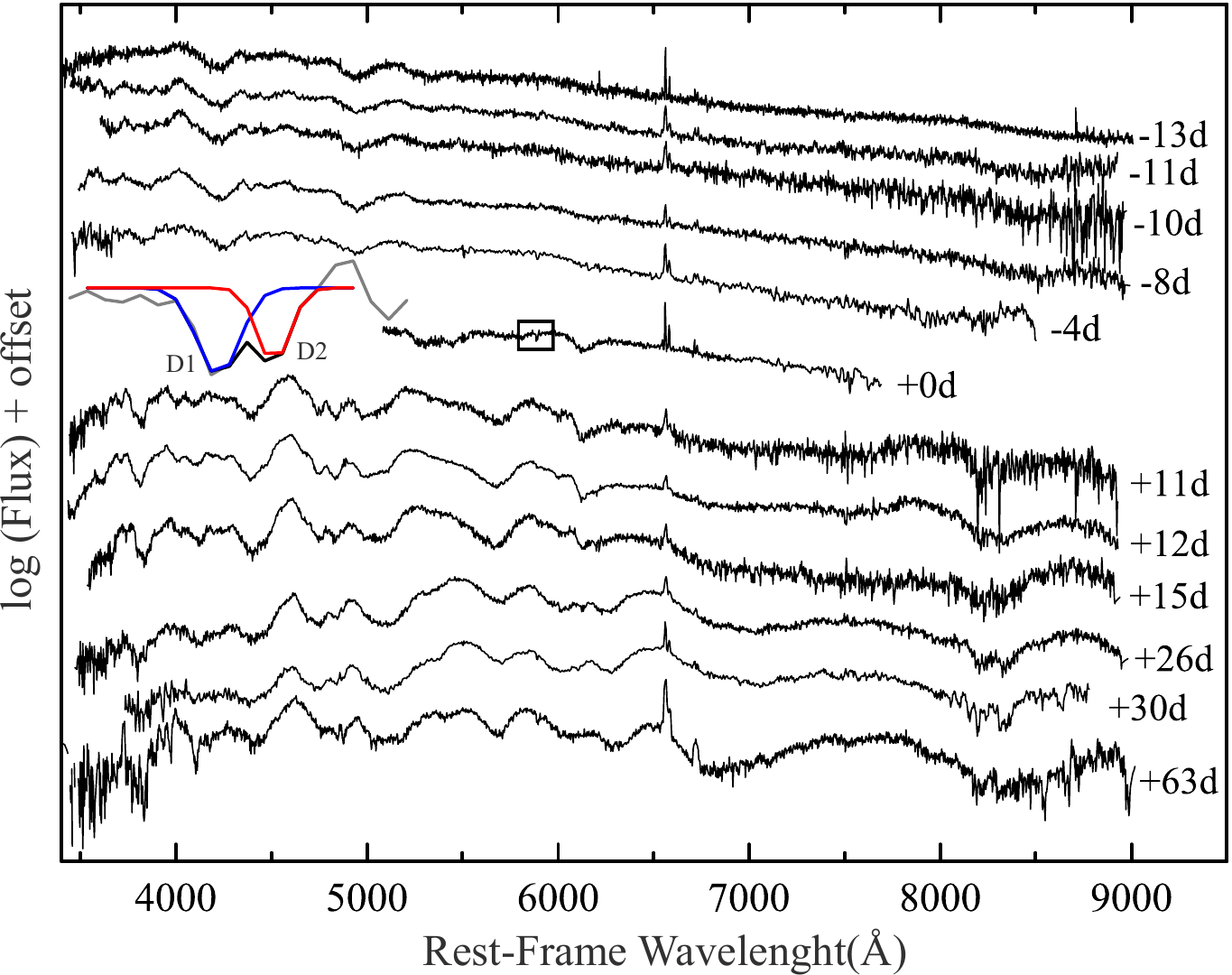}
 \caption{Optical spectral evolution of SN 2011hr. The spectra have been corrected for the redshift of the host galaxy ($v_{\rm hel}$ = 4137 \kms) and telluric lines. They have been shifted vertically by arbitrary amounts for clarity. The numbers on the right-hand side mark the epochs of the spectra in days since the $B$ band maximum. The \ion{Na}{1} D absorption of the host galaxy in the $t\approx+0$ day spectrum is indicated with a box, and the blow-up of this absorption is also shown on the left.}
\label{<Whsp>}
\end{figure}

A journal of spectroscopic observation of SN 2011hr is given in Table \ref{Tab:Spec_log}. A total of 11 low resolution spectra were obtained with the LJT and the Yunnan Faint Object Spectrograph and Camera (YFOSC, \citealp{JJzhang14}), and the 2.16-m telescope (hereinafter XLT) and the Beijing Faint Object Spectrograph and Camera (BFOSC). During our spectral observations, the typical FWHM was about $1\arcsec.5$ at Lijiang Observatory and about $2.\arcsec5$ at Xinglong Observatory.  All spectra were reduced using standard IRAF routines. Flux calibration was done using spectrophotometric flux standard stars observed at a similar air mass on the same night. The spectra were further corrected for the atmospheric absorption and telluric lines at the site. Synthetic photometry computed using \citet{Bess90} passbands was introduced to check the spectroscopic flux calibration, while the spectral fluxes were adjusted to match the contemporaneous photometry. Figure \ref{<Whsp>} shows the complete spectral evolution of SN 2011hr.

\begin{deluxetable*}{cccccccc}

\tablecaption{Journal of Spectroscopic Observations of SN 2011hr}

\tablehead{\colhead{Date} & \colhead{MJD} & \colhead{Epoch\tablenotemark{a}} & \colhead{Res.} & \colhead{Range} & \colhead{Slit Width} & \colhead{Exp. time} & \colhead{Telescope} \\
\colhead{} & \colhead{(-240000.5)} & \colhead{(days)} & \colhead{(\AA/pix)} & \colhead{(\AA)} & \colhead{(pix)} & \colhead{(sec)} & \colhead{(+Instrument)} }

\startdata
Nov. 10 2011    	&	55875.90		&	-13.22		&	1.5		&	 5000-9800 		&	2.8		&	1200		&	 LJT	YFSOC \\
Nov. 10 2011    	&	55875.92		&	-13.20		&	1.8		&	 3340-7680 		&	2.8		&	1200		&	 LJT	YFSOC \\
Nov. 12 2011 		&	55877.85		&	-11.27		&	2.9		&	 3500-8890 		&	4.2		&	1200		&	 LJT	YFSOC \\
Nov. 13 2011 		&	55878.85		&	-10.27		&	1.5		&	 5000-9800 		&	4.2		&	1200		&	 LJT	YFSOC \\
Nov. 13 2011 		&	55878.87		&	-10.25		&	1.8		&	 3340-7680 		&	4.2		&	1200		&	 LJT	YFSOC \\
Nov. 13 2011 		&	55878.89		&	-10.23		&	2.9		&	 3500-8890 		&	4.2		&	1200		&	 LJT	YFSOC \\
Nov. 15 2011 		&	55880.82		&	-8.30		&	2.9		&	 3500-8890 		&	4.2		&	1200		&	 LJT	YFSOC \\
Nov. 19 2011 		&	55884.84		&	-4.28		&	4.1		&	 3580-8565 		&	3.0		&	3600		&	 XLT	BFOSC \\
Nov. 23 2011 		&	55888.82		&	-0.30		&	2.1		&	 5080-7690 		&	3.0		&	1800		&	 XLT	BFOSC \\
Dec. 04 2011 		&	55899.91	        &	+10.79		&	2.9		&	 3500-8890 		&	6.4		&	900		&	 LJT	YFSOC \\
Dec. 05 2011 		&	55900.93		&	+11.81		&	2.9		&	 3500-8890 		&	6.4		&	900		&	 LJT	YFSOC \\
Dec. 08 2011 		&	55903.81		&	+14.69		&	2.9		&	 3500-8890 		&	6.4		&	900		&	 LJT	YFSOC \\
Dec. 19 2011 		&	55914.78		&	+25.66		&	2.9		&	 3500-8890 		&	6.4		&	900		&	 LJT	YFSOC \\
Dec. 23 2011 		&	55918.57		&	+29.45		&	4.1		&	 3780-8900 		&	6.0		&	2700		&	 XLT	BFOSC \\
Jan. 25 2012 		&	55951.76		&	+62.64		&	2.9		&	 3500-8890 		&	6.4		&	3000		&	 LJT	YFSOC

\enddata
\tablecomments{Journal of spectroscopic observations of SN 2011hr.}
\tablenotetext{a}{Relative to $B$ band maximum on Nov. 22.3 2011, JD. 2455889.62}
\label{Tab:Spec_log}
\end{deluxetable*}

\subsection{Temporal Evolution of the Spectra}
\label{sect:Spe_analy}

The spectra of SN 2011hr are described in more detail and compared to those of SNe 1991T, 1999aa, 2005cf, 2007if, and 2009dc at different epochs in this section, see Figures \ref{<Spear>}, \ref{<Spmax>} and \ref{<Sppost>}. All of these spectra have been corrected for redshift and reddening. The line identifications labeled in these figures are based on the modeling results for SN 1991T \citep{Paolo95}. The spectra of SN 2011hr have been corrected for the contamination of host-galaxy emission lines by using the spectrum taken at the site of SN 2011hr (with the LJT and YFOSC) at $\sim$\,500 days after the maximum light.

\subsubsection{The Early Phase}
\label{subsect:Spe_ear}

\begin{figure}
\centering
\includegraphics[width=8.5cm,angle=0]{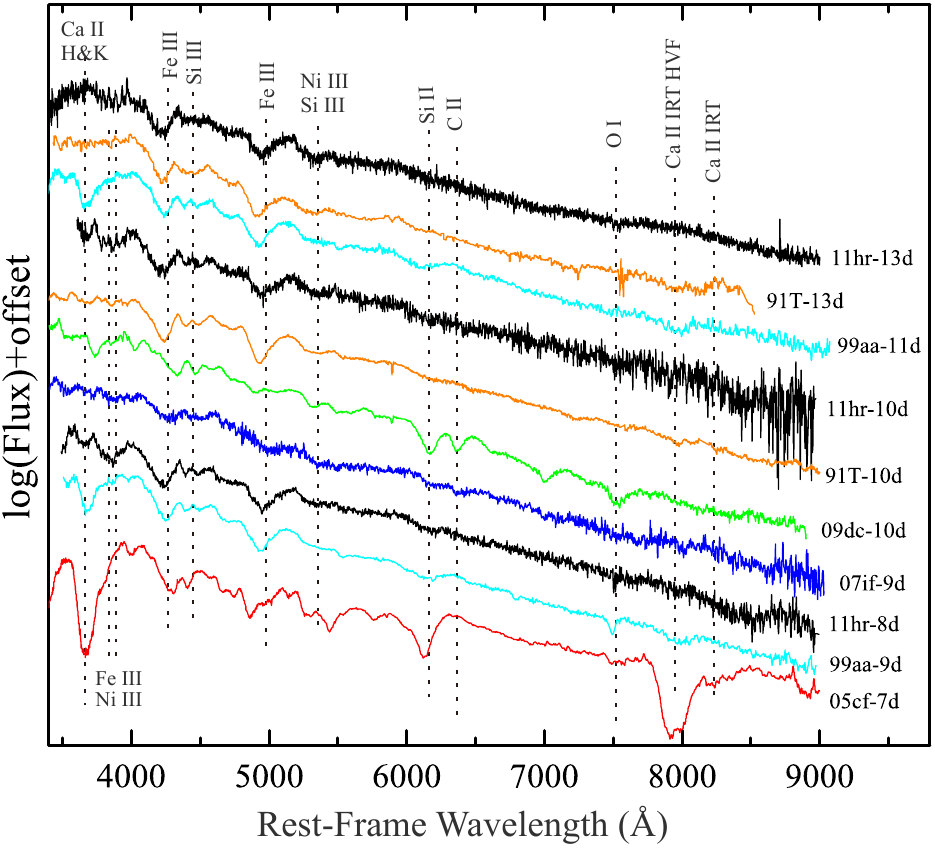}
 \caption{Spectra of SN 2011hr at $t < -7$ days before the $B$ band maximum light. Overplotted are the spectra of SNe 1991T \citep{Filip92a}, 1999aa \citep{99aa}, and 2005cf \citep{Wang09b} at comparable phases. The spectra of SN 2011hr are corrected for contaminations of the host galaxy.}
\label{<Spear>}
\end{figure}

\begin{figure}
\centering
\includegraphics[width=8.5cm,angle=0]{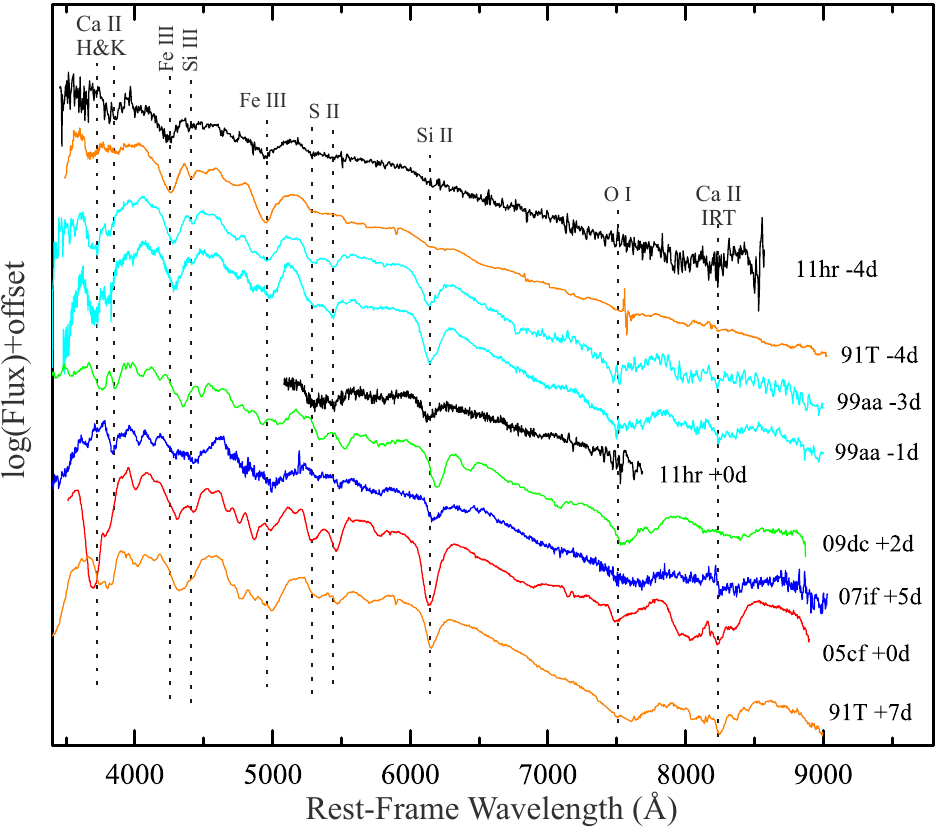}
 \caption{Spectra of SN 2011hr at around the maximum light, compared to those of the same SNe Ia as in Figure \ref{<Spear>}.}
\label{<Spmax>}
\end{figure}

\begin{figure}
\centering
\includegraphics[width=8.5cm,angle=0]{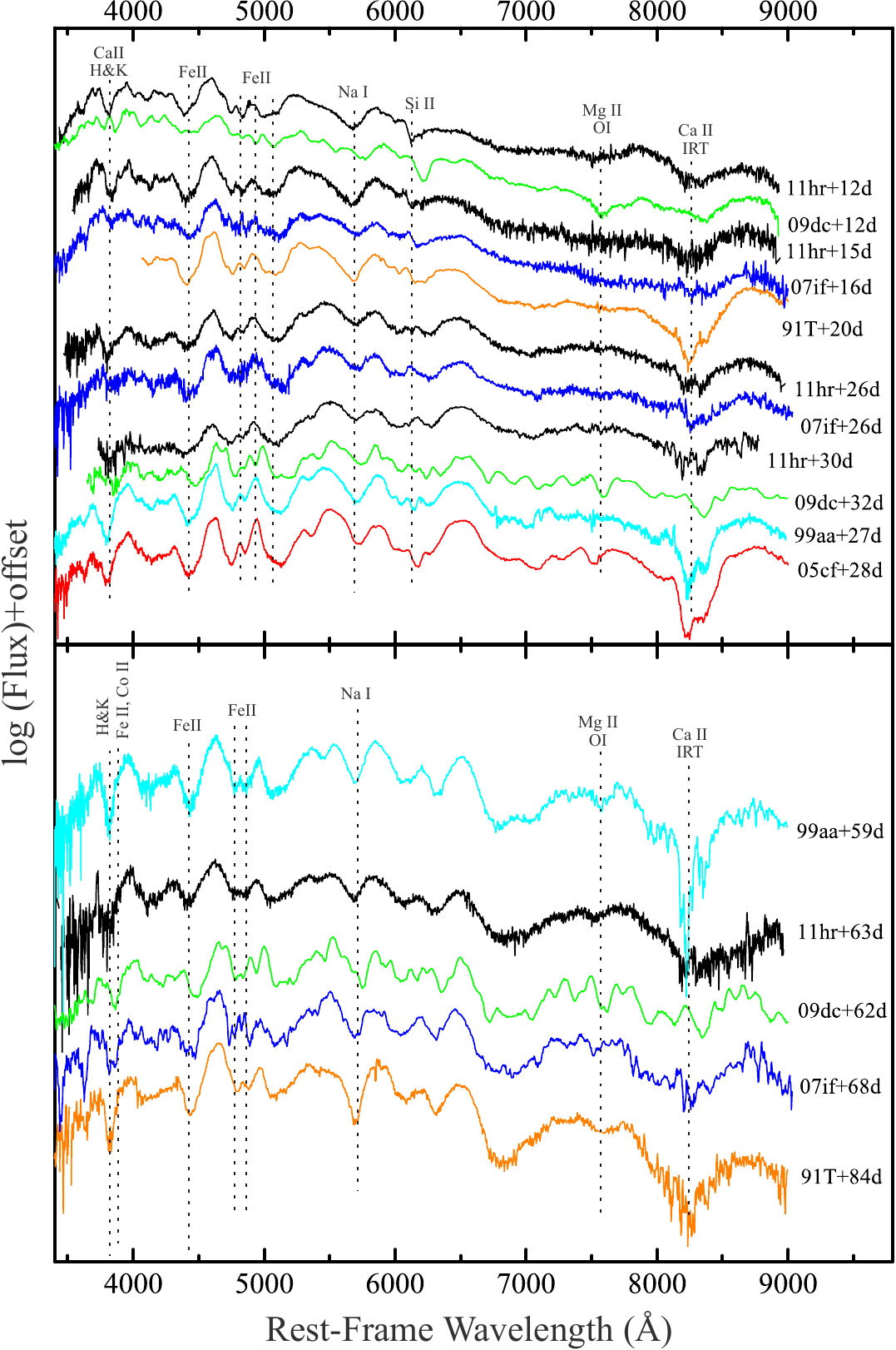}
 \caption{Post-maximum spectra of SN 2011hr, compared to the same SNe as in Figure \ref{<Spear>}. The spectrum of SN 2007if at $t\,\approx\,+68$ is smoothed with a box of 10 ($\sim$\,24\AA). The spectrum of SN 1999aa at $\,t\approx$\,+59 days is from \citet{Silver12}.}
\label{<Sppost>}
\end{figure}

Figure \ref{<Spear>} displays the spectra at $t < -7$ days. At this phase, the spectra usually carry signature of the ejecta materials in the outermost layers and show mainly lines of IMEs for normal SNe Ia. The spectra of SN 2011hr are rather similar to those of 91T-like events, which are characterized by a blue continuum and prominent \FeIII\,/\,\FeII~lines. Two main absorption features in SN 2011hr and SN 1991T are multiplets of \FeIII or \FeII lines at $\lambda$4404 and $\lambda$5129. By contrast, lines from IMEs (e.g., \SiII, S~{\sc ii} and O~{\sc i}) are very weak, which is a common characteristic of 91T-like SNe Ia. Lines of \SiIII\, are also present in the spectra. This peculiar phenomenon is due to a combination of high photospheric temperature and low abundance of light elements in the ejecta \citep{Paolo95,91T_14}.

\begin{figure}
\centering
\includegraphics[width=8.5cm,angle=0]{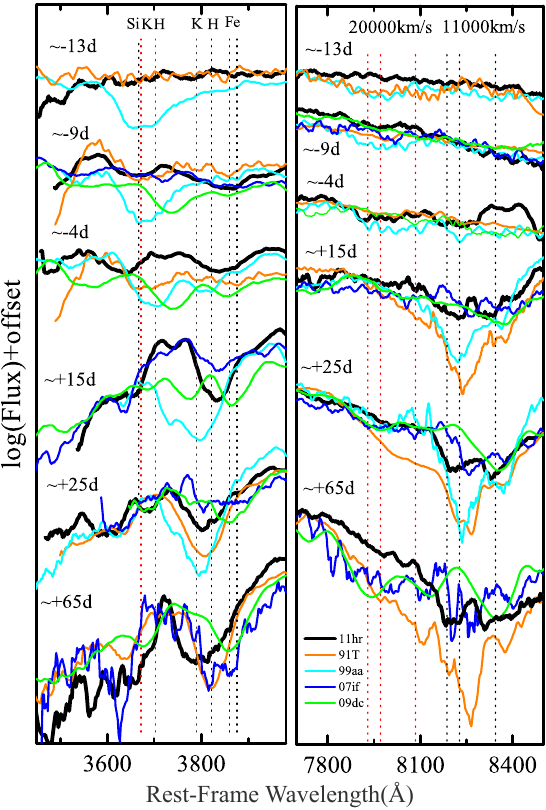}
 \caption{The evolution of the \CaII~absorption in the spectra of SNe 2011hr, 1991T, 1999aa, 2007if, and 2009dc at selected epochs. Left panels: The evolution of \CaII~H$\&$K and \SiII~$\lambda$3860. Right panels: The evolution of \CaII~NIR triplet. The positions of \CaII~H$\&$K, \ion{Si}{3}\,$\lambda$3807, \ion{Fe}{2} and the \CaII~NIR triplet lines at different velocities (black dotted-lines mark the line positions with $v$ = 11,000 \kms and red dotted-lines mark the positions with $v$ = 20,000 \kms) are shown to guide the eyes. All spectra are smoothed with a boxcar of 10 (i.e., $\sim$\,20\AA~--\,30\AA). See Figures \ref{<Spear>}, \ref{<Spmax>}, and \ref{<Sppost>} for the exact phase of each spectrum.}
\label{<CaII>}
\end{figure}

A notable difference between SN 2011hr and SN 1991T is the absorption near 3900\AA\ at $t\approx-10$ and $-8$ days. Such a feature could be a blend of \FeIII~$\lambda$4006 and \NiII~$\lambda$4067. The spectrum of SN 1999aa at $t\approx-9$ days shows both strong \FeIII\, lines and also a strong \CaII~H\,$\&$\,K feature, while the absorption of \SiII~$\lambda$6355 is weaker in this object. Spectral databases of SNe Ia (e.g., \citealp{Blondin12,Silver12}) indicate that most SNe show strong \CaII\ absorption from very early phases to 4\,--\,5 months after explosion. However, as shown in Figure \ref{<CaII>}, the \CaII\ lines in the spectra of SN 2011hr are even weaker than those of SN 1991T at some phases after the maximum light.

At $t < -7$ days, the spectrum of SN 2007if is also featureless and dominated by lines of doubly ionized Fe and Si. \CaII\ absorptions are also very weak at this phase. Note that the \FeIII lines are weaker in SN 2007if compared to those in SN 2011hr at this phase. In SN 2007if, a weak absorption on the red side of the weak \SiII~$\lambda$6355 could be due to absorption from \CII~$\lambda$6580. The spectrum of SN 2009dc at $t\approx -9$ days is characterized by prominent \CII~$\lambda$6580 absorption which is comparable in strength to the \SiII~$\lambda$6355. However, there is no evidence that such a strong \CII\ feature is present in the early-time spectra of SN 2011hr, which shows difference from the superluminous SNe~Ia.

\begin{figure}
\centering
\includegraphics[width=8.5cm,angle=0]{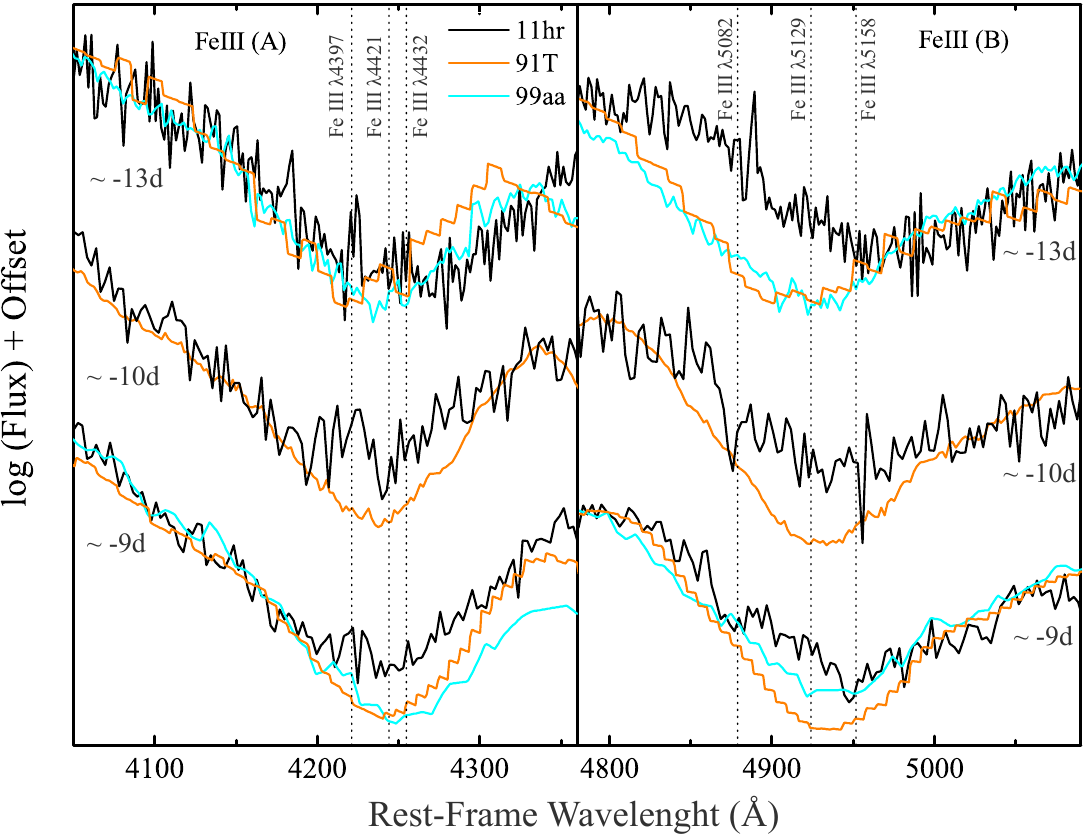}
 \caption{Absorption lines of \FeIII~in the spectra of SNe 2011hr, 1991T, and 1999aa at early phases. The positions of these lines at a velocity 12,000 \kms are marked to guide the eyes.}
\label{<Fe>}
\end{figure}

Figure \ref{<Fe>} compares the \FeIII~features of SN 2011hr, SN 1991T, and SN 1999aa. Some Fe lines, identified by modeling performed in Section \ref{subsect:Spmo}, are marked at a velocity of $\sim$\,12,000 km s$^{-1}$. The absorption feature shown in the left panels of the plot is a blend of \FeIII~$\lambda$4397, $\lambda$4421 and $\lambda$4432, and is represented as \FeIII~(A). The feature shown in the right panels is a multiplet of \FeIII~$\lambda$5082, $\lambda$5129, and $\lambda$5158, and is represented as \FeIII~(B). For the \FeIII~(A) feature, the three SNe Ia show similarities in the line strength, velocity, and shape at $t\approx-13$ days. For the \FeIII~(B) feature, SN 2011hr appears weaker than SN 1991T and SN 1999aa on the blue side of the absorption, which is more significant at earlier phases. It is unlikely that this difference is due to the absorption of \FeIII~$\lambda$5082 alone.

\subsubsection{The Maximum-light Phase}
\label{subsect:Spe_ear}
\begin{figure}
\centering
\includegraphics[width=8.5cm,angle=0]{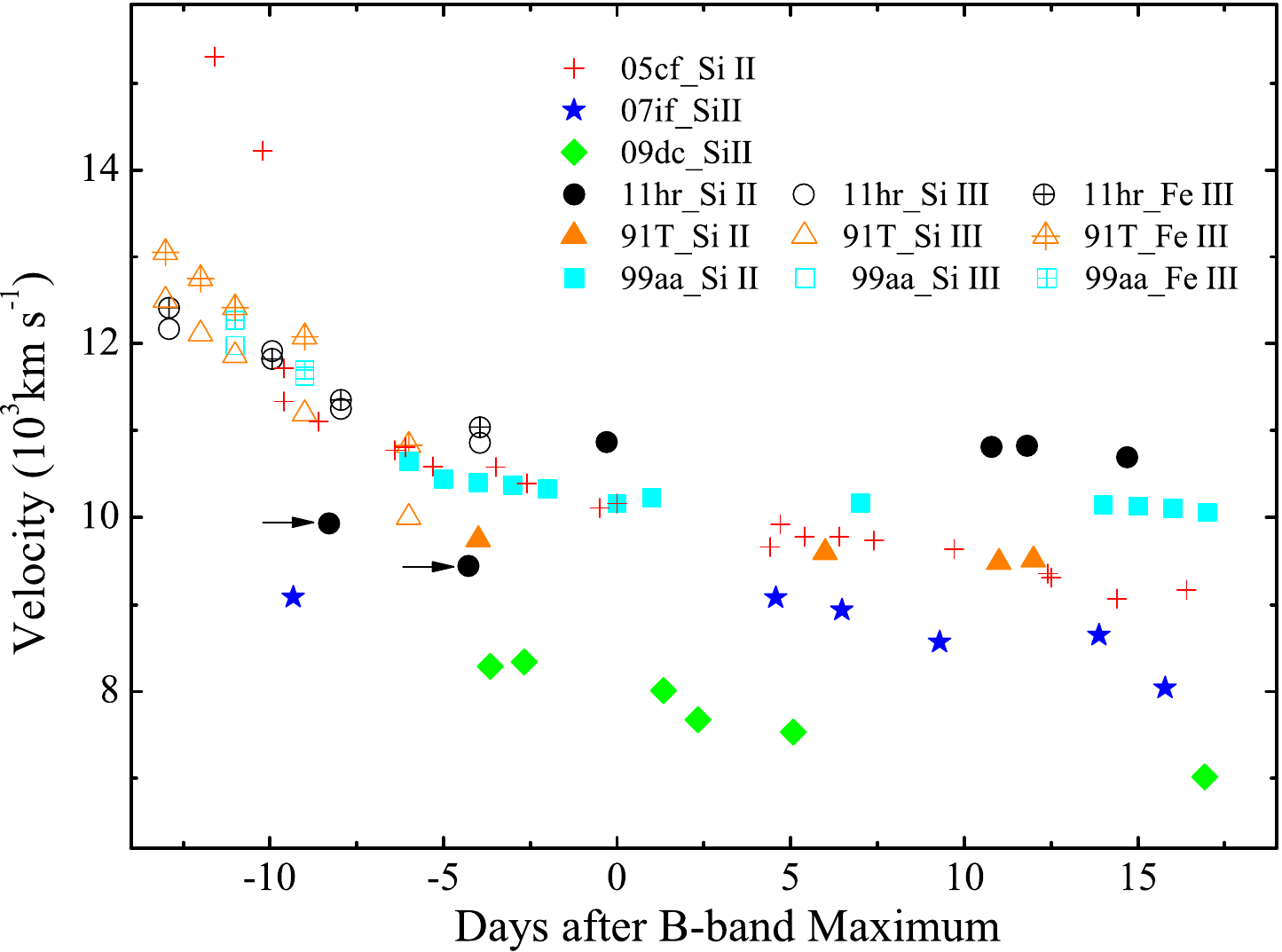}
 \caption{Ejecta velocity evolution of SNe 2011hr, 1991T, 1999aa, 2005cf, 2007if, and 2009dc, as inferred from absorption minima of \SiII\ $\lambda$6355, \SiIII\ $\lambda$4565, and \FeIII\ $\lambda$$\lambda$4400, 5100 lines.}
\label{<Vel>}
\end{figure}

Figure \ref{<Spmax>} shows the spectra at around the maximum light. At $t\approx -4$ days, the \FeII/\FeIII~lines appear shallower and wider in SN 2011hr compared to SN 1991T. At this phase, the S~{\sc ii}~features at $\sim$\,5300\AA~and $\sim$\,5450\AA~ start to emerge in the spectra of SN 2011hr and SN 1991T. These lines are useful to follow the shift of the ionization state of IMEs from doubly to singly ionized species. S~{\sc ii}~features in the spectra of SN 2011hr and SN 1991T are weaker than in SN 1999aa, suggesting a relatively lower ionization state for the latter. Moreover, the \FeIII~(B) absorption in SN 1999aa may be blended with the \FeII~lines, suggesting a lower ionization. In SN 2011hr, the weak absorption near 6150\AA~ is likely to be \SiII~$\lambda$6355, which may indicate a lower ejecta velocity (see also Figure \ref{<Vel>}). In other SNe~Ia, this \SiII~feature is stronger at similar phases.

At $t\,\approx\,0$ days, \SiII~$\lambda$6355 finally develops in SN 2011hr, but it does not become as strong as normal SNe Ia. The EW of this line is $\sim$\,34\AA~for SN 2011hr, which is only slightly larger than SN 2007if (i.e., $\sim$\,25\AA) and smaller than SN 1991T (i.e., $\sim$\,48\AA), SN 1999aa (i.e., $\sim$\,56\AA), SN 2005cf (i.e., $\sim$\,85\AA), and SN 2009dc (i.e., $\sim$\,51\AA) at similar phases. At this phase, the ``w"-shaped feature of S~{\sc ii} lines becomes noticeable in SN 2011hr. The presence of the S~{\sc ii}~lines indicates that SN 2011hr originated from a thermonuclear runaway. Inspection of Figure \ref{<Spmax>} reveals that there is no clear signature of O~{\sc i}~$\lambda$7773 absorption in SN 2011hr and SN 1991T near the maximum light. However, this feature becomes obvious in the spectra of other SNe~Ia at this phase.

\subsubsection{The Postmaximum Phase}
\label{subsect:Spe_post}

The postmaximum spectra of SN\,2011hr are compared with those of other SNe\,Ia in Fig. \ref{<Sppost>}. At about two weeks after the maximum light, the spectrum of SN 2011hr is dominated by lines of iron and sodium; and the \SiII\,$\lambda$6355 absorption is still very weak and \OI\,$\lambda$7773 line is almost undetectable at this phase. SN 2007if exhibits similar features in the comparable-phase spectra, while SN 2009dc shows stronger absorptions of \SiII\,$\lambda$6355 and \OI\,$\lambda$7773 but it has lower expansion velocities. Moreover, the \CaII~NIR triplet of SN 2011hr, SN 2007if, and SN 2009dc are all weak compared to SN 1991T.

At $t\,\approx\,+26$ days, the spectra of SN 2011hr and SN 2007if become more similar. For normal SNe~Ia at this phase, the absorption of \CaII~NIR triplet usually evolves to become the strongest feature in the spectra. However, such a line feature is still rather weak in SN 2011hr, SN 2007if, and SN 2009dc. At this phase, SN 1991T and SN 1999aa show absorption of \CaII~NIR triplet that is as strong as that seen in normal SNe Ia. At 2--3 months later, absorption of \CaII~NIR triplet grows in strength in SN 2011hr, SN 2007if, and SN 2009dc, but it is still significantly weaker compared to that seen in SN 1991T and SN 1999aa. The post-maximum spectra of SN 2011hr show distinct behavior relative to normal SNe Ia but are overall very similar to those of SN 2007if, including the velocity and strength of different species.

\subsection{Velocities of the Ejecta}
\label{subsect:Vel}

 m
The velocity gradient of SN 2011hr, derived from absorption of \SiII~$\lambda$6355 during the period from $t\,\approx\,+0$ to $t\,\approx$\,+15 days, is $9\,\pm10$\,\kms day$^{-1}$. This is comparable to that of SN 1999aa, SN 1991T, and SN 2007if, which places SN 2011hr in the low-velocity gradient (LVG) category of SNe Ia according to the classification scheme of \citet{Ben05}. After the maximum light,  SN\,2011hr shows a higher \SiII\ velocity relative to the comparison SNe\,Ia, indicating that the inner boundary of the Si shell is located at higher velocities. If the inner layers are dominated by Fe-group elements \citep{mazzali07}, this implies that SN\,2011hr produced more \Nifs\ than the normal SNe~Ia.

\section{Discussion}
\label{sect:Discu}band

\subsection{Quasi-bolometric Luminosity}
\label{subsect:bolo}
\begin{figure}
\centering
\includegraphics[width=8.5cm,angle=0]{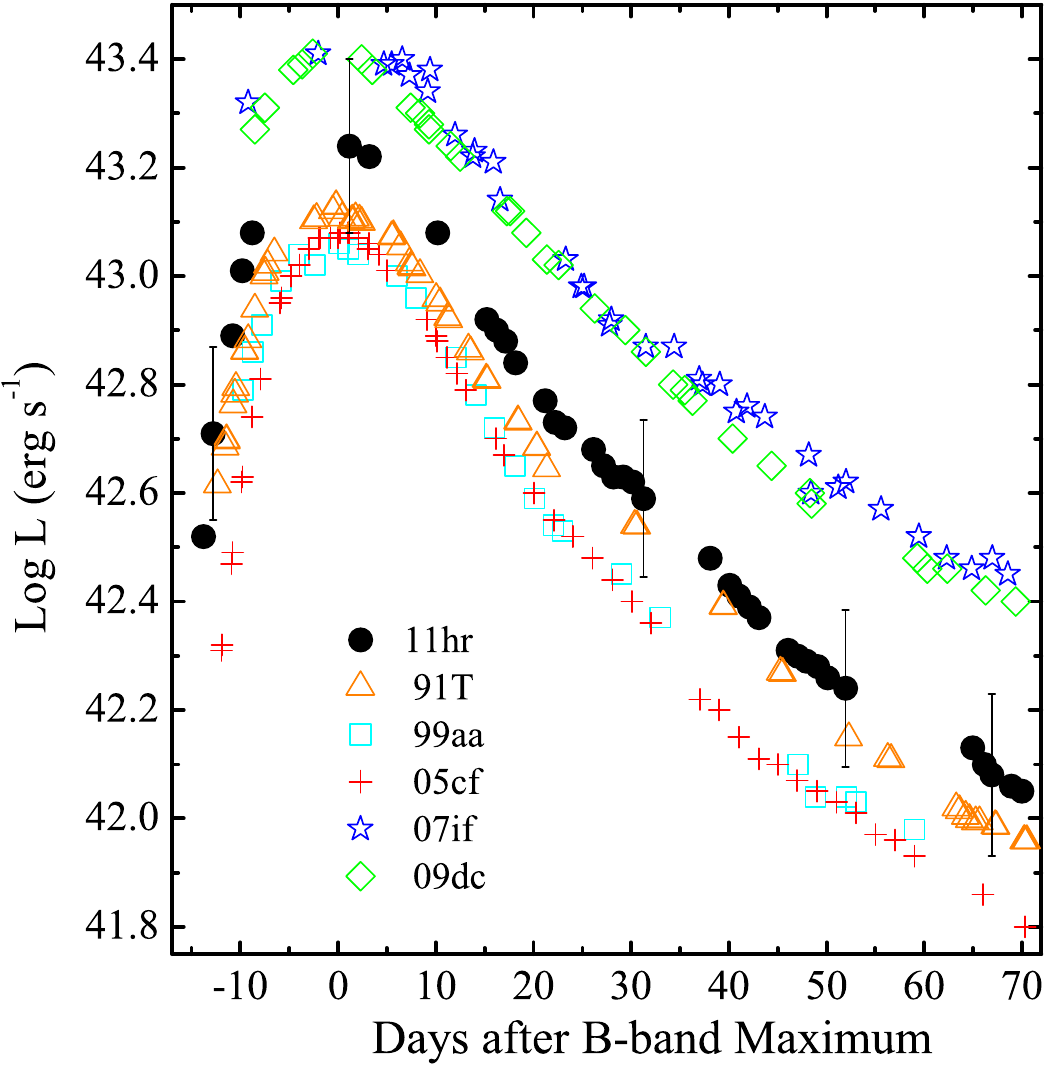}
 \caption{Quasi-bolometric light curves of SN 2011hr derived from the $UBVRI$ photometry, compared with those of SNe 1991T, 1999aa, 2005cf, 2007if and 2009dc.}
\label{<bolo>}
\end{figure}

Figure \ref{<bolo>} displays the quasi-bolometric light curves of SN 2011hr derived from the $UBVRI$ photometry, compared to those of SNe 1991T \citep{91T_14}, 1999aa \citep{Jha}, 2005cf \citep{Wang09b}, 2007if \citep{Scalzo07if}, and 2009dc \citep{Tau09dc}.
Note that the brightness of SN 2011hr locates at the middle of SN 1991T/SN 1999aa and SN 2007if/SN 2009dc. However, the bolometric light curve of SN 2011hr is similar to that of SN 1991T and SN 1999aa in morphology.  The superluminous SN 2007if and SN 2009dc show apparently wider light curves. These suggest that SN 2011hr may have a somewhat different explosion mechanism from these superluminous SNe Ia, although it shares some spectroscopic properties with SN 2007if.

To estimate the $^{56}$Ni mass synthesized in the explosion of SN 2011hr, we need to calculate its generic bolometric luminosity at maximum light. The UV emission of SN 2011hr was estimated by assuming that it has a similar UV to $B$ band band flux ratio as the well-sampled 99aa-like event iPTF14bdn \citep{14bdn}, while the NIR contribution was extrapolated based on its $J\&H$  bandband  photometry published by \citet{Anja11hr}. The peak ``$uvoir$" bolometric luminosity is thus estimated to be $2.30\,\pm\,0.90 \times 10^{43}$ erg s$^{-1}$. Applying a $t^{2}$ model to the observed $R$ band light curve yields an explosion time of MJD\,=\,55869.7\,$\pm$\,1.0 and a rise time of 19.4\,$\pm$\,1.0 days in $B$ band, which is close to the estimate derived from the spectral modeling (i.e., $\sim$\,20 days in the $B$ band; see discussions in \S 4.1). Considering the UV contribution, the bolometric curve of SN 2011hr reaches its peak at about 1.4 days earlier than in $B$ band; a rise time of 18.0 $\pm$ 1.0 day days is thus adopted for SN\,2011hr in the following analysis. Based on these parameters, we estimate that the $^{56}$Ni mass synthesized in the explosion of SN 2011hr is M(\Nifs)=$1.11\,\pm\,0.43 {\rm M}_{\sun}$ according to the Arnett law \citep{Arn82,56Ni05}. This value  is larger than that obtained for SN\,1991T (M(\Nifs) = 0.82 M$_{\sun}$), SN\,1999aa (M(\Nifs) = 0.72 M$_{\sun}$), and SN 2005cf (M($^{56}$Ni) = 0.75 M$_{\sun}$). The $^{56}$Ni masses inferred for SN\,2007if and SN 2009dc (with a similar method) are much larger, which are close to 2.0 M$_{\sun}$. An accurate determination of the \Nifs\ mass could be also obtained using nebular spectra \citep{mazzali07}, but no late-time spectra is available for SN\,2011hr.

\subsection{Explosion models}
\label{subsect:model}

Our observations indicate that SN~2011hr is an overluminous SN Ia. Both the super-Chandrasekhar (SC)-mass and the Chandrasekhar (Ch)-mass scenario have been proposed to explain observational features of overluminous SNe Ia. \citet{Yoon05} suggested that the WDs may accrete mass from a non-degenerate or another WD companion star and grow in mass to values exceeding the Chandrasekhar-mass limit (which can reach a mass of $\gtrsim2.0\,M_{\odot}$ if the WD is in differential rotation). Thermonuclear explosions of rapidly rotating WDs have been suggested to account for production of superluminous SNe Ia in terms of both burning products and explosion kinematics by \citet{Pfannes10a,Pfannes10b}. Also, they find that a rotating WD that detonates has an ejecta velocity compatible with that of a normal SN Ia. Meanwhile, \citet{Pfannes10a,Pfannes10b} also showed that a significant amount of IMEs is produced for rotating SC-mass WDs. Inspection of Figure~\ref{<Spmax>} reveals that SN 2011hr has rather weak absorption features of Si~II, \CaII, and O~I, unlike the 91T-like and normal SNe~Ia. These results indicate that the IMEs (especially Ca) are apparently less abundant in SN 2011hr, which seems to be inconsistent with the predictions of rotating SC-mass WDs (see \citealt{Pfannes10a,Pfannes10b}).

Alternatively, SN~2011hr may be an energetic explosion of a Chandrasekhar-mass WD. For instance, the Chandrasekhar-mass model with a high kinetic energy in \citet{WDD3} may be a good candidate for producing overluminous SNe Ia, which will be addressed below in Section~\ref{subsect:Spmo}. The gravitationally confined detonation (GCD) of near-Chandrasekhar-mass WDs may be an alternative scenario to explain the overluminous SNe~Ia. The GCD model has been studied by different groups (e.g., \citealp{Plewa04,Plewa07,Meakin09,Jordan12}). On one hand, it has been shown that the GCD model may provide explanations for overluminous SN~1991T-like events because it can produce a large mass of $^{56}\rm{Ni}$ that can even exceed 1.0\,M$_{\odot}$. On the other hand, GCD models naturally produce a chemically mixed compositions and thus are proposed to explain the high-velocity Fe-group elements observed in early-time spectra of some overluminous 91T-like SNe. Unfortunately, to date, no detailed radiative transfer calculations have yet been performed to compare with the observations of overluminous SNe Ia.

Finally, we point out that the absence of C and O lines in the early-time spectra of SN~2011hr suggests that SN~2011hr is unlikely to be the result of the merger of two WDs (i.e., the so-called violent merger model within the double-degenerate scenario).

\subsection{Abundance tomography of SN~2011hr}
\label{subsect:Spmo}

To set further constraints on the possible progenitor scenario, we modeled the early spectra of SN~2011hr using a well tested Monte Carlo code that was initially developed by \citet{Paolo93} and later improved by \citet{Lucy99} and \citet{Paolo00} by including the photon branching. The code assumes a photosphere with a scaled black-body emission. This approximation is valid when the inner part of the ejecta that encloses most of the radioactive materials is optically thick. To model the spectra we adopt a technique called ``abundance tomography'' \citep{Stehle,Paolo08,Tanaka11,Ashall14,91T_14}. As shown in previous sections, there are some uncertainties in determinations of the distance and host-galaxy reddening for SN~2011hr. A higher luminosity may require a SC-mass progenitor, e.g., M(\Nifs) = 1.61 M$_{\sun}$ if $E(B-V)\,=\,0.27$ and D\,=\,66 Mpc. We want to investigate whether a Chandrasekhar-mass scenario is still viable for SN~2011hr, or whether it can only be explained with an SC-mass model. For our investigation, we use the density profile of the energetic WDD3 model from \citet{WDD3}. This should be a good choice for SN 2011hr, which shows lines with relatively higher velocities.

\begin{figure}
\begin{center}
\includegraphics[ angle=0, width=0.48\textwidth]{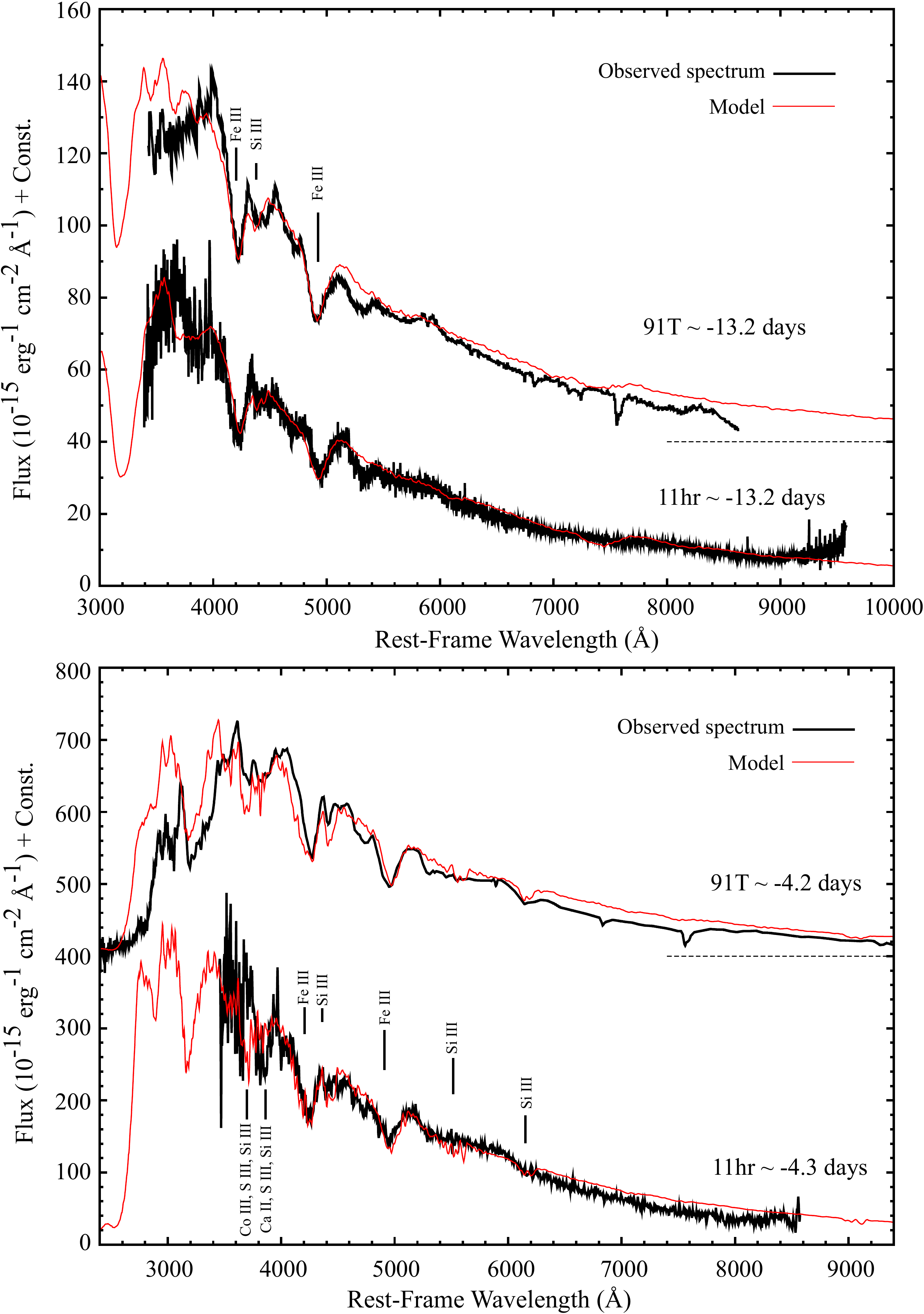}
\caption{Synthetic spectra are compared with the observed spectra of SN\,2011hr at $-13.2$ and $-4.3$ days.  At the top of each panel, the model and the observed spectrum of SN\,1991T at the same epoch from \citet{91T_14} are shown for comparison. To allow the comparison between these two SNe, the flux of SN\,2011hr has been scaled by putting it at the same distance as SN\,1991T. The spectra are in rest-frame and corrected for reddening by the amount assumed during the modeling.}
 \label{N13}
\end{center}
\end{figure}

\begin{figure*}
\begin{center}
\includegraphics[angle=0, width=0.8\textwidth]{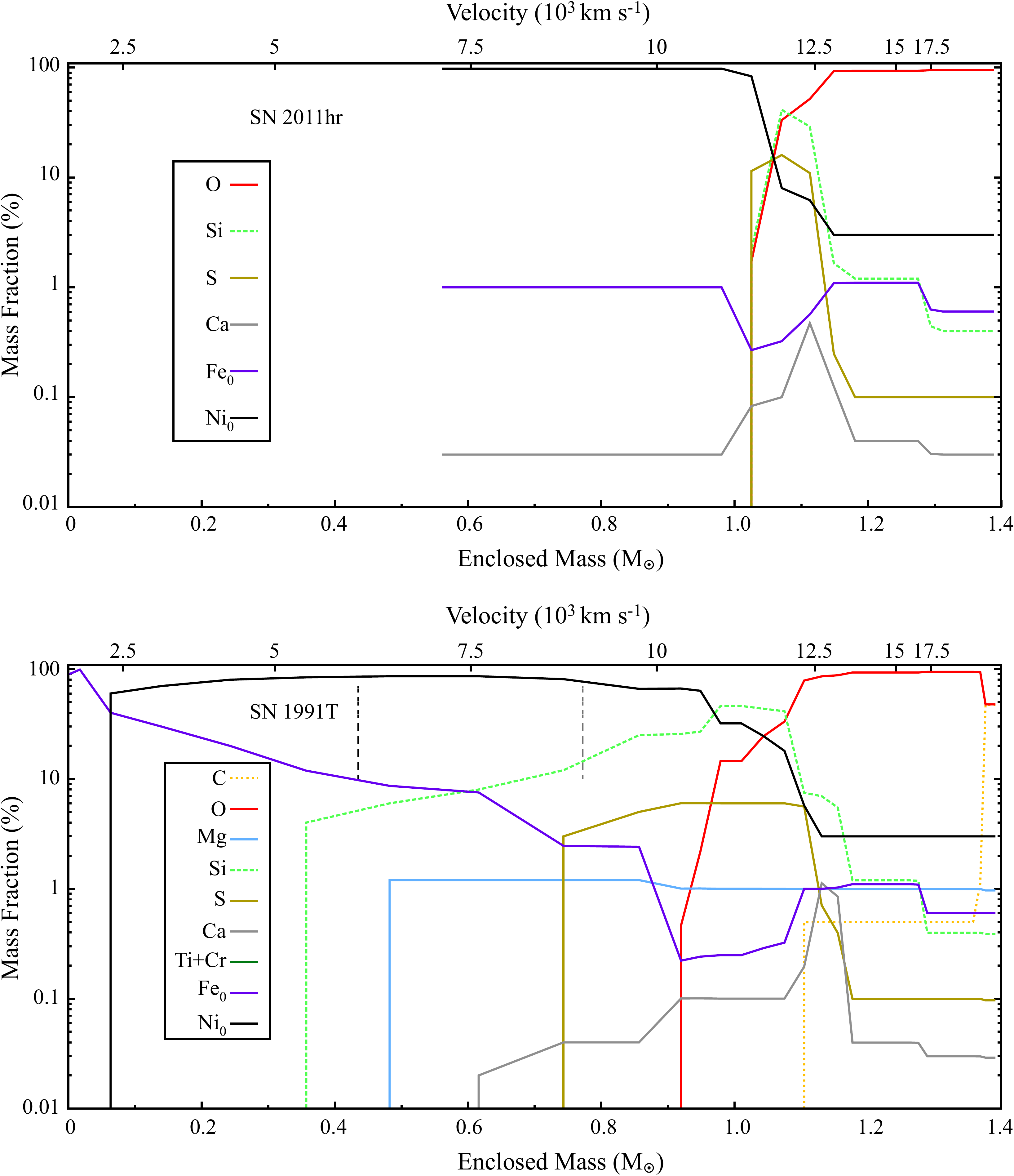}
\caption{The abundance distribution of the outer ejecta inferred for SN\,2011hr and SN\,1991T. SN\,2011hr (top panel) has a thinner layer of IMEs above a large amount of \Nifs. Only the outer layers can be probed with the photospheric method adopted here. The abundances of SN\,1991T (bottom panel) are taken from Fig.5 of \citet{91T_14}.}
 \label{abundances}
\end{center}
\end{figure*}

We fit the luminosity of SN\,2011hr from spectral modeling. This is an independent estimate of the distance and the extinction. We follow the same method as that described in \citet{91T_14} to study the luminosity of SN 1991T, which is an ideal SN for a comparative study with SN~2011hr. Similarly to SN~1991T, the time at which the recombination from \SiIII\ into \SiII\ happens can place a tight constraint on the luminosity. For SN 2011hr, this transition happens at a couple of days later than SN~1991T. Figure \ref{N13} displays the comparison of the modelled spectra of SN~2011hr at $t\approx -13$ days and $-4$ days, respectively. The observed and modelled spectra of SN~1991T are also shown as references. The distance adopted for SN\,2011hr is $D=60$ Mpc (i.e., $\text{m}-\text{M}\,=\,33.89$). In addition to the Milky-Way reddening of $E(B-V) = 0.02$, we use a reddening $E(B-V) = 0.15$ for the host galaxy. A larger reddening is disfavored by the synthetic color from the model spectrum.

Compared to SN~1991T, the \FeIII\ features at 4200\,\AA\ and 4900\,\AA\ are less pronounced in SN\,2011hr. The modeling analysis shows that this can be explained with an ionization effect, as a consequence of the fact that the temperature of SN\,2011hr is higher than that of SN~1991T. The \SiII\ line at 6100\,\AA\ first appears at around the $B$ band maximum light, $\sim4$ days later than in SN~1991T. Again, this can be also due to an ionization effect caused by the higher luminosity of SN~2011hr.  On the other hand, the \SiIII\ 4450\AA\ line is weaker than in SN\,1991T despite a larger fraction of \SiIII\ because of a lower Si abundance. Si is the dominant element only within 11500 and 12000 \kms. SN~2011hr has an abundance profile qualitatively similar to SN~1991T, but likely representing a more extreme case (i.e., Figure \ref{abundances}). The interface between the \Nifs-dominated interior and the layer dominated by IMEs is thinner in SN 2011hr compared to that seen in SN 1991T. The \Nifs\ fraction exceeds 90\% up to velocities as high as $\sim11500$\,\kms. This matches the lowest velocities of \SiII\ measured in the spectra. The layer dominated by Si and other IMEs is also thinner than in SN~1991T, and their total mass is just $0.07\,\Msun$, while in SN\,1991T the IME mass is $0.18\,\Msun$ \citep{91T_14}. We could recap the properties of SN\,2011hr as ``more 91T-like than SN~1991T itself''.

The modeling result shows that the luminosity of SN~2011hr is higher than SN~1991T by about 20\%. This requires a \Nifs\ mass of about $0.94\ \Msun$. The luminosity of SN\,2011hr is at the brightest end of SNe\,Ia, which requires a large amount of \Nifs, but this is balanced by a lower mass of IMEs. The model does not require a mass larger than the Chandrasekhar mass, and instead it suggests that a scenario with an efficiently burned Chandrasekhar mass is also possible to produce such an overluminous SN Ia.

\section{Conclusion}
\label{sect:con}

We have presented extensive observations and studies of the luminous (e.g., $M_{\rm max}$(B) = $-19.84 \pm 0.40$ mag), spectroscopically peculiar type Ia supernova SN 2011hr. It exhibits light curve evolution that is similar to SN 1991T, with a post2.4 maximum decline rate $\Delta m_{15}$(B)=0.92\,$\pm$\,0.03 mag. The spectra of SN 2011hr are characterized by strong Fe-group lines and weak absorptions of IMEs, showing close resemblances to that of SN 2007if at all phases. Note, however, that the absence of \CII~ absorption in the early-time spectra of SN 2011hr suggests that there should be difference between these two SNe Ia. Moreover, SN\,2011hr has narrower light curve peaks and its light-curves decline more rapidly at late phases compared to SN 2007if. Overall, SN 2011hr shows more similarities with SN 1991T in morphology of the light curves, while it bears more common features with SN 2007if in the spectral evolution.

Computing the detailed abundance distribution of the IMEs and iron-group elements helps further reveal the nature of this peculiar SN Ia. Mapping the abundances in the outer ejecta with the abundance tomography method we found that a lower mass of IMEs (i.e., 0.07 $\Msun$) and a larger mass of \Nifs\  (i.e., $\sim$\ 0.94 $\Msun$ by spectral modeling or $\sim$\ 1.11 $\Msun$ by Arnett law) can be inferred for SN 2011hr. The corresponding values inferred for SN\,1991T is $\sim$ 0.18 $\Msun$ for IMEs and $\sim$0.78 $\Msun$ for \Nifs. Our models suggest that a very efficiently burned Chandrasekhar-mass progenitor is the favoured scenario for this overluminous SN Ia. The weaker absorptions of the IMEs might indicate a lower abundance of these elements in the outer ejecta. Nevertheless, the presence of similar features in SN\,2011hr and SN\,2007if may indicate that SN\,2011hr represents a transitional object linking 91T-like SNe Ia to some superluminous SNe\,Ia.

\acknowledgments

We acknowledge the support of the staff of the Li-Jiang 2.4 m telescope (LJT), Xing-Long 2.16 m telescope, and Tsinghua-NAOC 0.8 m telescope (TNT). Funding for the LJT has been provided by the Chinese Academy of Sciences (CAS) and the People's Government of Yunnan Province. The TNT is owned by Tsinghua University and jointly operated by the National Astronomical Observatory (NAOC) of the CAS.   Financial support for this work has been provided by the National Science Foundation of China (NSFC, grants 11403096, 11178003, 11325313, 11133006, 11361140347, 11203034, 11473063, 11322327, 11103072, 11203078, 11303085, 11203070 and 11103078); the Major State Basic Research Development Program (2013CB834903); the Strategic Priority Research Program ``The Emergence of Cosmological Structures" of the CAS (grant No. XDB09000000); the Key Research Program of the CAS (Grant NO. KJZD-EW-M06);  the Western Light Youth Project; the Youth Innovation Promotion Association of the CAS; the Open Project Program of the Key Laboratory of Optical Astronomy, NAOC, CAS; and the key Laboratory for Research in Galaxies and Cosmology of the CAS.

\clearpage


\begin{thebibliography}{}
\bibitem[Altavilla et al.(2004)]{Altavilla}Altavilla, G., Fiorentino, G., Marconi, G. et al., 2004, MNRAS, 349, 1344
\bibitem[Arnett(1982)]{Arn82} Arnett, W. D. 1982, \apj, 253, 785
\bibitem[Ashall et al.(2014)]{Ashall14}Ashall C., Mazzali P., Bersier D., Hachinger S., Phillips M., Percival S., James P., Maguire K., 2014, MNRAS, 445, 4427
\bibitem[Barbon et al.(1990)]{Bar25}Barbon R., Benetti S., Rosino L., Cappellaro E., Turatto M., 1990, \aap, 237, 79
\bibitem[Barone-Nugent et al.(2012)]{Barone-Nugent12}Barone-Nugent, R. L., Lidman, C., Wyithe, J. S. B., et al. 2012, MNRAS, 425, 1007
\bibitem[Benetti(2005)]{Ben05} Benetti, S. 2005, \apj, 623, 1011
\bibitem[Bessell(1990)]{Bess90}Bessell, M. S. 1990, PASP, 102, 1181
\bibitem[Blondin \& Tonry (2007)]{SNID} Blondin,S., \& J.L., Tonry. 2007, \apj, 666, 1024
\bibitem[Blondin et al.(2012)]{Blondin12} Blondin,S.,   Matheson,T.,  Kirshner, R. P., et al. 2012, \aj, 143, 126
\bibitem[Branch et al.(2005)]{Branch05} Branch, D., Baron, E., Hall, N.,  et al. 2005, \pasp, 117, 545
\bibitem[Branch et al.(2006)]{Branch06} Branch, D.,  Dang, L. C., Hall, N., et al. 2006, \pasp, 118, 560
\bibitem[Branch et al.(2009)]{Branch09} Branch, D., Dang, L.C., $\&$ Baron, E. 2009, \pasp, 121, 238
\bibitem[Brown et al.(2014)]{BrownUV}Brown, P.J., Kuin, P., Scalzo, R., et al. 2014, \apj, 787, 29
\bibitem[Cardelli et al.(1998)]{Cardelli} Cardelli, J. A., Clayton, G. C., $\&$ Mathis, J. S. 1989, \apj, 345, 245
\bibitem[Childress et al.(2013)]{C13}Childress, M.J., Scalzo, R.A., Sim, S., et al. 2013, \apj, 770, 29
\bibitem[Childress et al.(2014)]{ChildHVF} Childress, M.J.,Filippenko, A.V., Ganeshalingam, M., et al. 2014, \mnras, 437, 338
\bibitem[Cousins (1981)]{Cousins}Cousins, A. 1981, South African Astron. Obs. Circ., 6, 4
\bibitem[Dilday et al.(2012)]{11kx} Dilday, B., Howell, D.A., Cenko, S.B., et al. 2012, Science, 337, 942
\bibitem[Fan et al.(2015)]{FYF} Fan, Y.F., Bai, J.M., Zhang, J.J., et al. 2015, Research in Astron. Astrophys,  15, 918
\bibitem[Filippenko et al.(1992a)]{Filip92a} Filippenko, A.V.,  Richmond, M.W.,  Michael, W., et al. 1992a, \apj, 384, L15
\bibitem[Filippenko et al.(1992b)]{Filip92b} Filippenko, A.V., Richmond, Michael W., Branch, David., et al. 1992b, \aj, 104, 1543
\bibitem[Filippenko(1997)]{Filip97}Filippenko, A.V. 1997, \araa, 35, 309
\bibitem[Fisher et al.(1999)]{Fisher99}Fisher, A., Branch, D., Hatano, K., $\&$ Baron, E. 1999, MNRAS, 304, 67
\bibitem[Fitzpatrick \& Massa (2007)]{Fitzpatrick07}Fitzpatrick, E. L. \& Massa, D. 2007, \apj, 663, 320
\bibitem[Foley et al.(2013)]{Foley13} Foley, R. J., Challis, P.J., Chornock, R.,  et~al. 2013, \apj, 767, 57
\bibitem[Garavini et al.(2004)]{99aa}Garavini, G., Folatelli, G., Goobar, A., et al. 2004, \apj, 128, 387
\bibitem[Gibson $\&$ Stetson (2001)]{Gibson}Gibson, B. K., Stetson, P. B., 2001, ApJ, 547, L103
\bibitem[Goldhaber et al.(2001)]{Goldhaber01}Goldhaber, G.; Groom, D. E.; Kim, A., et al. 2001, \apj,  558, 359
\bibitem[Guy et al.(2005)]{Guy05}Guy, J.; Astier, P.; Nobili, S., et al. 2005, \aap,  443, 781
\bibitem[Hachinger et al.(2012)]{Hach09dc} Hachinger S., Mazzali P.A., Taubenberger S., et al. 2012, \mnras, 427, 2057
\bibitem[Hamuy et al.(1996)]{Hamuy96} Hamuy M., Phillips M. M., Suntzeff N. B., et al. 1996, \aj,  12, 2438
\bibitem[Hicken et al. (2007)]{Hicken07}Hicken M., Garnavich P.M., Prieto J.L., et al. 2007, ApJ, 669, L17
\bibitem[Hicken et al.(2009)]{Hicken09}Hicken M., Wood-Vasey, W. M., Blondin, S.,  et al., 2009, \apj, 700, 1097
\bibitem[Hillebrandt et al.(2007)]{Hillebrandt07}Hillebrandt W., Sim S.A., \& R\"{o}pke F.K. 2007, \aap, 465, 17
\bibitem[Howell et al.(2006)]{Howell06}Howell D.A.,  Sullivan, M., Nugent, P. E., et al. 2006, Nature, 443, 308
\bibitem[Huang et al.(2012)]{TNT}Huang, F., Li, J.Z., Wang, X.F., et al. 2012, Research in Astron. Astrophys, 11, 1585
\bibitem[Iwamoto et al.(1999)]{WDD3} Iwamoto, K., Brachwitz, F., Nomoto K., et al. 1999, ApJS, 125, 439
\bibitem[Johnson et al.(1966)]{Johnson}Johnson, H., Iriarte, B., Mitchell, R., $\&$ Wisniewskj, W. 1966, Commun. Lunar Planet. Lab., 4, 99
\bibitem[Jordan et al.(2012)]{Jordan12}Jordan IV, G.C.,  Graziani C., Fisher R.T., et al. \apj, 2012, 759, 53
\bibitem[Jha et al.(2006)]{Jha}Jha, S., Krishner, R.P., Challis, P., et al. 2006, \aj, 131, 527
\bibitem[Krisciunas et al.(2004)]{Kri04}Krisciunas, K., Phillips, M. M., \& Suntzeff, N. B. 2004, ApJ, 602, L81
\bibitem[Landolt (1992)]{Landolt}Landolt, A. V. 1992, AJ, 104, 340
\bibitem[Li et al.(2003)]{Li02cx}Li, W., Filippenko, A., Ryan, C., et al. 2003, \pasp, 115, 453
\bibitem[Lira et al.(1998)]{Lira98}Lira, P., Suntzeff, Nicholas B., Phillips, M. M., et al. 1998, \aj, 115, 234
\bibitem[Lucy (1999)]{Lucy99}Lucy L. B. 1999, \aap, 345, 211
\bibitem[Maeda et al.(2010)]{Maeda10}Maeda K., Benetti, S., Stritzinger, M., et al. 2010, Nature, 466, 82
\bibitem[Maeda et al.(2011)]{Maeda11}Maeda K.,  Leloudas G., Taubenberger  S.,  et al. 2011, \mnras, 413, 3075
\bibitem[Maoz et al.(2014)]{Maoz14} Maoz, D., Mannucci, F., $\&$ Nelemans, G. 2014, \araa, 52, 107
\bibitem[Marion et al. (2013)]{marion13} Marion, G. H., Vinko, J., Wheeler, J.C, et al. 2013, \apj, 777, 40
\bibitem[Maund et al.(2010)]{Maund10}Maund, J., H\"{o}rflich, P., Patat, F.,  et al. 2010, \apj, 725L, 167
\bibitem[Mazzali \& Lucy (1993)]{Paolo93}Mazzali P. A., Lucy L. B., 1993, A\&A, 279, 447
\bibitem[Mazzali et al.(1995)]{Paolo95}Mazzali, P.A., Danziger, I.J., \& Turatto, M., 1995 \aap, 297, 509
\bibitem[Mazzali (2000)]{Paolo00}Mazzali, P.A. 2000, \aap, 363, 705
\bibitem[Mazzali et al.(2005)]{Paolo05}Mazzali, P. A.   Benetti, S., Altavilla, G., et al. 2005,  ApJL, 623, L37
\bibitem[Mazzali et al.(2007)]{mazzali07} Mazzali, P.~A., R{\"o}pke, F.~K., Benetti, S., \& Hillebrandt, W.\ 2007, Science, 315, 825
\bibitem[Mazzali et al.(2008)]{Paolo08}Mazzali, P. A., Sauer, D. N., Pastorello, A., Benetti, S., Hillebrandt, W., 2008, MNRAS, 386, 1897
\bibitem[Meakin et al.(2009)]{Meakin09}Meakin C.A., Seitenzahl I., Townsley D., et al. 2009, \apj, 693, 1188
\bibitem[Meikle (2000)]{Meik00}Meikle, W. P. S. 2000, MNRAS, 314, 782
\bibitem[Mould et al.(2000)]{Mould00}Mould, J.R., Huchra, J. P., Freedman, W. L., 2000, ApJ, 529, 786
\bibitem[Nayak et al.(2011)]{CBET2901a} Nayak, I., Cenko, S.B., Filippenko, A.V., 2011, CBET 2901, 1
\bibitem[O'Donnell (1994)]{ODonnell94}O'Donnell, J. E. 1994, \apj, 422, 158
\bibitem[Perlmutter et al.(1999)]{Perl99} Perlmutter, S.,  Aldering, G., Goldhaber, G., et al. 1999, \apj, 517, 565
\bibitem[Pfannes et al.(2010a)]{Pfannes10a}Pfannes J.M.M., Niemeyer J.C., Schmidt W., Klingenberg C. Thermonuclear explosions of rapidly rotating white dwarfs. I. Deflagrations. 2010a, \aap, 509, 74
\bibitem[Pfannes et al.(2010b)]{Pfannes10b}Pfannes J.M.M., Niemeyer J.C., \& Schmidt W. Thermonuclear explosions of rapidly rotating white dwarfs. II. Detonations. 2010b, \aap, 509, 75
\bibitem[Phillips et al.(1992)]{Phillip92} Phillips, M. M., Wells, L. A., Suntzeff, N. B., et al. 1992, \aj, 103, 1632
\bibitem[Phillips(1993)]{Phillip93} Phillips, M. 1993, \apj, 413, L105
\bibitem[Phillips et al.(1999)]{Phillip99} Phillips, M., Lira, R., Suntzeff. N.B., et al. 1999, \aj, 118, 1766
\bibitem[Phillips et al.(2013)]{Phillip2013}Phillips, M., Simon, J.,  Morrell, N., et al. 2013, \apj, 779, 38
\bibitem[Plewa et al.(2004)]{Plewa04}Plewa T., Calder A.C., Lamb D.Q., et al. 2004, \apj, 612, L37
\bibitem[Plewa et al.(2007)]{Plewa07}Plewa T.  2007, \apj, 657, 942
\bibitem[Poznanski et al.(2011)]{Ext11}Poznanski, D., Ganeshalingam, M., Silverman, J.M., $\&$ Filippenko, A.V. 2011, \mnras, 415, L81
\bibitem[Riess et al.(1996)]{Riess96}Riess, A., Press, W., and Kirshner, R. 1996, \apj,  473, 88
\bibitem[Riess et al.(1998)]{Riess98} Riess, A., Filippenko, A.V., Challis, P., et al. 1998, \aj, 116, 1009
\bibitem[Ruiz-Lapuente et al.(1992)]{Ruiz92}Ruiz-Lapuente, P., Cappellaro, E., Turatto, M., et al. 1992, \apj, 387L, 33
\bibitem[Sasdelli et al.(2014)]{91T_14}Sasdelli, M., Mazzali, P.A., Pian, E., et al. 2014, \mnras, 445, 711
\bibitem[Scalzo et al.(2010)]{Scalzo07if} Scalzo, R., Aldering, G., Antilogus, P., et al. 2010, \apj, 713, 1073
\bibitem[Schmidt et al.(1998)]{Schmidt98}Schmidt, B., Suntzeff, N., Phillips, M., et al. 1998, \apj, 507, 46
\bibitem[Schlegel et al.(1998)]{Schle98}Schlegel, D., Finkbeiner, D., and Davis, M., 1998, \apj, 500, 525
\bibitem[Silverman et al.(2011)]{Silver11} Silverman, J.,  Ganeshalingam, M.,  Li, W., et al.  2011, \mnras, 410, 585
\bibitem[Silverman et al.(2012)]{Silver12} Silverman, J.,  Foley, R.J.,  Filippenko, A.V., et al.  2012, \mnras, 425, 1789
\bibitem[Sim et al.(2007)]{Sim07}Sim, S. Sauer, D. N., R\"{o}pke, F. K., Hillebrandt, W.  2007, MNRAS, 378, 2
\bibitem[Smitka et al.(2015)]{14bdn}Smitka, M. T., Brown, P.J.,  Suntzeff, N.B., et al. 2015, \apj, 813, 30
\bibitem[Stehle et al.(2005)]{Stehle}Stehle, M., Mazzali, P. A., Benetti, S., Hillebrandt, W., 2005, MNRAS, 360, 1231
\bibitem[Stetson (1987)]{Stetson}Stetson, P. 1987, \pasp, 99, 191
\bibitem[Stritzinger \& Leibundgut(2005)]{56Ni05}Stritzinger, M., \& Leibundgut, B. 2005, \aap, 431, 423
\bibitem[Suntzeff et al.(1996)]{Sun96}Suntzeff, N. 1996, in Supernovae and Supernova Remnants, ed. McCray, R.,  $\&$ Wang, Z. (Cambridge:Cambridge Univ.Press), p41
\bibitem[Tully et al.(2009)]{Tully}Tully, R. B., Rizzi, L., Shaya, E., et al. 2009, AJ, 138, 323
\bibitem[Tanaka et al.(2011)]{Tanaka11}Tanaka M., Mazzali P. A., Stanishev V., Maurer I., Kerzendorf W. E., Nomoto K., 2011, MNRAS, 410, 1725
\bibitem[Taubenberger et al.(2011)]{Tau09dc}Taubenberger, S.,  Benetti, S.,  Childress, M., et al. 2010, \mnras, 412, 2735
\bibitem[Theureau et al.(2007)]{DisTF}Theureau, G., Hanski, M. O., Coudreau, N., et al. 2007, \aap, 465, 71
\bibitem[Turatto et al. (2003)]{Tura16}Turatto, M., Benetti, S.,  \& Cappellaro, E., 2003, in From Twilight to Highlight: The Physics of Supernovae, Hillebrandt, W.,  Leibundgut, B., eds., p.200
\bibitem[Wang et al.(2008)]{wang08}Wang, X., Li, W., Filippenko, A.V., et al. 2008, \apj, 675, 626
\bibitem[Wang et al.(2009a)]{Wang09a} Wang, X., Filippenko, A.V., Ganeshalingam, M., et al.  2009a, \apj, 699, L139
\bibitem[Wang et al.(2009b)]{Wang09b} Wang, X., Li, W., Filippenko, A., et al.  2009b, \apj, 697, 380
\bibitem[Wang et al.(2012)]{04dtUV}Wang, X.F., Wang, L.F.,  Filippenko, A.V, et al. 2012, \apj, 749, 126
\bibitem[Wang et al.(2013)]{Wang13} Wang, X., Wang, L., Filippenko, A., et al.  2013, Science, 340, 170
\bibitem[Weyant et al.(2014)]{Anja11hr}Weyant, A., Michael, W., Allen, L., et al. 2014, \apj, 784, 105
\bibitem[Yoon \& Langer (2005)]{Yoon05}Yoon S.C., \& Langer N. 2005, \aap, 435, 967
\bibitem[Zhang et al.(2011)]{CBET2901b} Zhang, T.M., Zhang, J.J., $\&$ Wang, X.F. 2011, CBET 2901, 2
\bibitem[Zhang et al.(2014)]{JJzhang14} Zhang, J.J, Wang, X.F., Bai, J.M., et al., 2014, \aj, 148, 1
\end{thebibliography}
\end{document}